\documentclass[pdflatex,sn-mathphys-num, iicol]{sn-jnl}% Math and Physical Sciences Numbered Reference Style
\usepackage{graphicx}%
\usepackage{multirow}%
\usepackage{amsmath,amssymb,amsfonts}%
\usepackage{amsthm}%
\usepackage{mathrsfs}%
\usepackage[title]{appendix}%
\usepackage{xcolor}%
\usepackage{textcomp}%
\usepackage{manyfoot}%
\usepackage{booktabs}%
\usepackage{algorithm}%
\usepackage{algorithmicx}%
\usepackage{algpseudocode}%
\usepackage{listings}%
\usepackage{soul}
\theoremstyle{thmstyleone}%
\theoremstyle{thmstyletwo}%

\theoremstyle{thmstylethree}%

\begin{document}

\title[Article Title]{Indirect Reciprocity with Environmental Feedback}


\author[1,4,6]{\fnm{Yishen} \sur{Jiang}}
\author*[2,4,6,7,9]{\fnm{Xin} \sur{Wang}}\email{wangxin\_1993@buaa.edu.cn}
\author[1,4,6]{\fnm{Ming} \sur{Wei}}
\author[2,4,6]{\fnm{Wenqiang} \sur{Zhu}}
\author[2,4,6,7]{\fnm{Longzhao} \sur{Liu}}
\author[10]{\fnm{Hongwei} \sur{Zheng}}
\author*[2,3,4,5,6,7,8]{\fnm{Shaoting} \sur{Tang}}\email{tangshaoting@buaa.edu.cn}

\affil[1]{School of Mathematical Sciences, Beihang University, Beijing 100191, China}
\affil[2]{School of Artificial Intelligence, Beihang University, Beijing 100191, China}
\affil[3]{Hangzhou International Innovation Institute, Beihang University, Hangzhou 311115, China}
\affil[4]{Key Laboratory of Mathematics, Informatics and Behavioral Semantics, Beihang University, Beijing 100191, China}
\affil[5]{Institute of Medical Artificial Intelligence, Binzhou Medical University, Yantai 264003, China}
\affil[6]{Zhongguancun Laboratory, Beijing 100094, China}
\affil[7]{Beijing Advanced Innovation Center for Future Blockchain and Privacy Computing, Beihang University, Beijing 100191, China}
\affil[8]{Institute of Trustworthy Artificial Intelligence, Zhejiang Normal University, Hangzhou, 310013}
\affil[9]{State Key Laboratory of General Artificial Intelligence, BIGAI, Beijing, China}
\affil[10]{Beijing Academy of Blockchain and Edge Computing, Beijing 100085, China}


\abstract{Indirect reciprocity maintains cooperation in stranger societies by mapping individual behaviors onto reputation signals via social norms. Existing theoretical frameworks assume static environments with constant resources and fixed payoff structures. However, in real-world systems, individuals’ strategic behaviors not only shape their reputation but also induce collective-level resource changes in ecological, economic, or other external environments, which in turn reshape the incentives governing future individual actions. To overcome this limitation, we establish a co-evolutionary framework that couples moral assessment, strategy updating, and environmental dynamics, allowing the payoff structure to dynamically adjust in response to the ecological consequences of collective actions. We find that this environmental feedback mechanism helps lower the threshold for the emergence of cooperation, enabling the system to spontaneously transition from a low-cooperation state to a stable high-cooperation regime, thereby reducing the dependence on specific initial conditions. Furthermore, while lenient norms demonstrate adaptability in static environments, norms with strict discrimination are shown to be crucial for curbing opportunism and maintaining evolutionary resilience in dynamic settings. Our results reveal the evolutionary dynamics of coupled systems involving reputation institutions and environmental constraints, offering a new theoretical perspective for understanding collective cooperation and social governance in complex environments.}




\maketitle

\section{Introduction}\label{sec1}

The emergence of large-scale cooperation among non-kin individuals remains a classic puzzle in biology and the social sciences~\cite{west2007evolutionary, perc2017statistical}. In typical pairwise interactions, individuals incur costs to provide benefits to others. Without external constraints, such unidirectional altruism is vulnerable to exploitation by selfish free-riders, leading to the collapse of cooperation~\cite{hardin1968tragedy, rankin2007tragedy, zhu2025evolution, meng2025promoting}. Particularly in stranger societies characterized by high mobility and a lack of repeated interactions, direct reciprocity mechanisms often fail to function effectively~\cite{boyd1988evolution, fehr2003nature, nowak2006five}. To address this dilemma, indirect reciprocity has been proposed as a mechanism based on reputation and social norms, with the core idea that people are more inclined to cooperate with those of good social standing~\cite{trivers1971evolution, nowak1998evolution}. Unlike direct reciprocity, it extends interactions to third parties: individuals help others not for immediate returns from the recipient, but to accumulate a positive social reputation~\cite{kandori1992social, tomasello2013origins, fehr2018normative, schmid2021unified}. In both psychology and sociology, this mechanism has been shown to sustain stable cooperative orders within broad social networks~\cite{ostrom1990governing, tomasello2013origins, bird2015prosocial, curry2019good, von2019dynamics}.

The effectiveness of indirect reciprocity relies on social norms, the assessment rules by which individual behaviors are mapped onto reputation signals. Early pioneering work identified the ``Leading Eight" norms, establishing a benchmark for second- and higher-order norms to maintain stable cooperation~\cite{ohtsuki2004should, ohtsuki2006leading}. Building on this, more recent research has shifted from simple norm comparison to analyzing how norms influence cooperation through reputation channels under information and cognitive constraints~\cite{ohtsuki2009indirect, uchida2010effect, wei2025indirect}. Reputation-conditional evolutionary game models indicate that when strategies and reputations change over time, the emergence and maintenance of cooperation depend sensitively on how the reputation system operates~\cite{milinski2002reputation, nowak2005evolution, sigmund2010calculus, zhu2024reputation}. In complex information environments, the mode of reputation generation and transmission determines the boundaries of cooperation: when reputation is widely shared through institutional records or gossip, individuals can reach a consensus, supporting higher levels of cooperation~\cite{fehr2004third, gurerk2006competitive, sommerfeld2007gossip, radzvilavicius2021adherence, kessinger2023evolution}; conversely, private assessments and stereotypes lead to reputation disagreement~\cite{hilbe2018indirect, kawakatsu2024stereotypes, schmid2023quantitative}, while perceptual noise and systemic bias blur reputation judgments~\cite{ohtsuki2009indirect, righi2022gossip, kawakatsu2024mechanistic}, all of which may drive cooperative arrangements to break down~\cite{fujimoto2023evolutionary}. Recent research further shows that in the absence of public monitoring, appropriate tolerance policies can help eliminate subjective disagreement and stabilize cooperation~\cite{michel2024evolution}.

Parallel to the scrutiny of internal assessment rules, evolutionary game theory has also explored the critical role of the external ecological context in the emergence of cooperation. Real-world interactions invariably occur under the constraints of fluctuating resource states, where environmental factors reshape selection pressures through multiple dimensions~\cite{roca2009evolutionary, cressman2014replicator}. Static environmental heterogeneity fundamentally alters payoff structures, thereby delineating the intensity of social dilemmas~\cite{zhu2024evolutionary}. Environmental noise arising from seasonal fluctuations or stochastic perturbations has been shown to divert evolutionary trajectories, potentially inducing resonance effects under periodic variations~\cite{gokhale2016eco,  taitelbaum2023evolutionary}. The framework of ``eco-evolutionary game theory" further elucidates that in many natural and social systems, the environmental state itself is subject to feedback from population behavior~\cite{weitz2016oscillating, wang2020eco}. On one hand, environmental states capable of stochastic switching based on behavior can significantly promote cooperation~\cite{ginsberg2018evolution, su2019evolutionary, gao2024evolutionary}. On the other hand, the bidirectional coupling between environment and strategies can induce complex dynamics, ranging from ``oscillating tragedies of the commons" to chaos~\cite{weitz2016oscillating, shao2019evolutionary, wang2020steering, tilman2020evolutionary, liu2022general, hua2024coevolutionary, jiang2025nonlinear}. Such coupled dynamics can be extended to spatial networks and ecological tipping points, revealing how local environmental feedback and non-linear threshold effects reshape the level and stability of cooperation~\cite{szolnoki2018environmental, jiang2023nonlinear, betz2024evolutionary}.

Despite the macro-dynamic perspective provided by eco-evolutionary games, the environmental dimension remains largely neglected within the framework of indirect reciprocity. Most existing models are built upon the donation game, assuming fixed payoff matrices and implicitly unlimited resources, examining how norms evaluate behavior via reputation only within highly simplified static backgrounds~\cite{nowak2005evolution, sasaki2017evolution, okada2018solution}. However, empirical research indicates that environment, strategy, and reputation are tightly intertwined in real systems~\cite{ostrom1993coping, rustagi2010conditional}. Howe et al. showed that in the presence of high environmental risk, resource sharing depends strongly on past reputation, functioning as a form of social insurance~\cite{howe2016indirect}. Milinski et al. found that reputation incentives effectively sustain high levels of cooperation when climate risks are distinct and contributions are visible~\cite{milinski2006stabilizing}. In addition, favorable economic environments tend to lead to more lenient social reputation evaluation standards, thereby promoting the emergence of cooperation~\cite{wei2025indirect}. Thus, reputation is not an abstract label independent of the environment, but an endogenous institution that dynamically adjusts with environmental pressure and strategies. This naturally leads to a question that has not yet been systematically answered: when the environmental dimension is incorporated into the indirect reciprocity framework, how does this tripartite coupling mechanism reshape the cooperative landscape and the relative fitness of social norms?

To address this issue, we construct a co-evolutionary framework that unifies strategy evolution, reputation assessment, and environmental dynamics, wherein the payoff structure evolves endogenously with the resource state. We find that under environmental feedback, discriminating strategies and strict norms can secure their dominance regardless of initial conditions, eliminating the bistability observed in static models and forming a 'locking effect' that robustly steers the population toward a stable state of cooperation. Furthermore, we uncover a ``paradox of tolerance" from a dynamic perspective: while lenient norms exhibit adaptability in static, resource-poor environments, only norms with strict discrimination can remain evolutionarily viable and sustain environmental resilience in dynamic settings. Our work not only establishes the patterns of strategy evolution and the hierarchy of norms under dynamic constraints but also highlights that endogenizing the environment is pivotal for understanding how human societies co-evolve with their resources to escape the tragedy of the commons.

\begin{figure*}[t]
\centering
\includegraphics[width=0.95\textwidth]{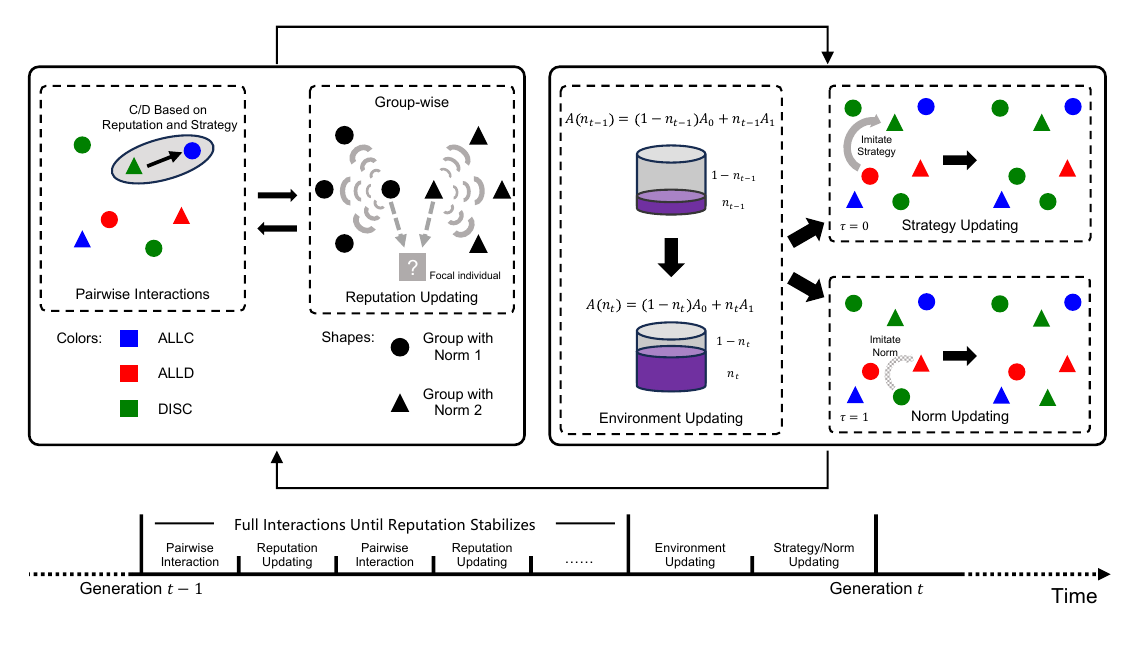}
\caption{\textbf{Schematic framework of the coevolutionary model of strategies, norms, and environment.} The timeline at the bottom illustrates the nested dynamics of the model: within each generation interval, repeated individual interactions occur first, followed by system-level intergenerational updates. In the intra-generational interaction phase (left boxes), individuals engage in pairwise games based on their own strategies and the reputation of their opponents; these actions are then evaluated according to the norms of their respective groups, leading to reputation updates. This "game-reputation" loop repeats until the reputation of the population reaches a quasi-steady state. The collective outcomes of these interactions drive changes in the environmental state, thereby altering the payoff context for subsequent games. At the end of a generation, based on accumulated fitness, the population undergoes strategy updates or norm updates through imitation mechanisms, determining the state of the next generation.}\label{fig1}
\end{figure*}

\section{Model}\label{sec2}

We construct an eco-evolutionary framework that integrates strategy evolution, reputation assessment, and environmental dynamics into a unified system. As illustrated in Fig.~\ref{fig1}, the model captures the interplay between microscopic interactions and macroscopic states through a nested loop structure. Within each generation, individuals engage in pairwise interactions where payoffs are modulated by the current environmental state, while reputations are updated according to specific social norms of groups. The collective outcome of these interactions then drives the feedback mechanism that reshapes the environment. Finally, based on accumulated fitness, the population undergoes evolutionary updates of strategies or norms, closing the adaptive loop.

\subsection{Pairwise interactions}

Consider a large, well-mixed population where individuals are randomly matched to play a one-shot interaction. In the classic setting of indirect reciprocity, the interaction is a donation game: when the donor cooperates, the recipient receives a benefit $b$ and the donor pays a cost $c$ with $b>c>0$; when the donor defects, no benefit is given and no cost is paid (we set $c=1$ in this paper). The matrix form $A_{1}= \begin{pmatrix} b & -c\\ b-c & 0 \end{pmatrix}$ is a special case of the Prisoner’s Dilemma. However, such a static payoff structure fails to capture the fluctuating nature of real socio-ecological systems where incentives are intrinsically tied to resource availability. This dependency could be well illustrated by the dynamics of a fishery. Abundant fish stocks lower the perceived cost of over-exploitation and encourage individuals to pursue immediate gains at the expense of the collective. Conversely, severe scarcity transforms collective restraint into a survival necessity and naturally suppresses the temptation to defect. To mathematically capture this shift, we introduce an environmental state variable $n \in [0,1]$ to interpolate between these two distinct strategic regimes. Specifically, the limit $n \to 1$ corresponds to the abundant state modeled by $A_1$, while the limit $n \to 0$ represents the scarce state governed by a matrix $A_{0}= \begin{pmatrix} b-c & 0\\ b & -c \end{pmatrix}$. Here, the prohibitive cost of mutual defection renders cooperation the rational choice to avert disaster. The environment-dependent payoff matrix is then defined as
\begin{align}
\begin{split}
A(n)&=(1-n)A_{0}+nA_{1} \\
&=(1-n)\begin{pmatrix} b-c & 0\\ b & -c \end{pmatrix} +n\begin{pmatrix} b & -c\\ b-c & 0 \end{pmatrix}.
\end{split}
\end{align}

In each match, the donor’s decision depends on its current strategy and the perceived reputation of the recipient. We consider three strategies: always cooperate (ALLC), always defect (ALLD), and a discriminator (DISC) that cooperates with a “good” recipient and defects otherwise. We include a cooperation execution error: if an individual intends to cooperate, the action flips to defection with probability $u_{c}$ with $0<u_{c}\ll 1/2$; intended defection is error-free.
Given these assumptions and the joint effects of environment and reputation, the expected payoffs of ALLC, ALLD, and DISC for group $i$ are
\begin{align}
\begin{aligned}
\pi_{i}^{\mathrm{ALLC}}
&=(1-u_{c})\Bigg[
  b\sum_{j}\nu_{j}\Big(f_{j}^{\mathrm{ALLC}}+f_{j}^{\mathrm{DISC}}\,r_{j,i}^{\mathrm{ALLC}}\Big)\\
&\qquad
  -n c
\Bigg],\\
\pi_{i}^{\mathrm{ALLD}}
&=(1-u_{c})\Bigg[
  b\sum_{j}\nu_{j}\Big(f_{j}^{\mathrm{ALLC}}+f_{j}^{\mathrm{DISC}}\,r_{j,i}^{\mathrm{ALLD}}\Big)\\
&\qquad
  -(1-n)c
\Bigg],\\
\pi_{i}^{\mathrm{DISC}}
&=(1-u_{c})\Bigg[
  b\sum_{j}\nu_{j}\Big(f_{j}^{\mathrm{ALLC}}+f_{j}^{\mathrm{DISC}}\,r_{j,i}^{\mathrm{DISC}}\Big) \\
&\qquad
  -n r_{i,\cdot}c-(1-n)\big(1-r_{i,\cdot}\big)c
\Bigg].
\end{aligned}
\end{align}
We partition the population into $K$ disjoint groups. The weight of group $j$ is $\nu_{j}$ with $\sum_{j=1}^{K}\nu_{j}=1$. Here $\pi_{i}^{S}$ is the expected payoff of strategy $S\in\{\mathrm{ALLC},\mathrm{ALLD},\mathrm{DISC}\}$ in group $i$, and $f_{j}^{S}$ is the fraction of group $j$ using $S$. The term $r_{j,i}^{S}$ is the probability that group $j$ evaluates a strategy-$S$ individual from group $i$ as “good.” The average reputation that group $i$ assigns to the whole population is
$r_{i,\cdot}=\sum_{l} \nu_{l}\,r_{i,l}$, where the term $r_{i,l}=\sum_{S} f_{l}^{S}\,r_{i,l}^{S}$ is the average reputation of group $l$ from the perspective of group $i$. 

\subsection{Reputations updating}

After each round of pairwise interaction, reputations are updated in a group-wise manner. In each group, one member is randomly chosen as an observer. This observer watches a focal individual acting as a donor paired with a randomly matched recipient. The observer evaluates the focal individual by combining the group’s current view of the recipient’s reputation with the focal individual’s action in that interaction. We adopt four second-order social norms. All of them agree that cooperating with a good recipient yields a good reputation and defecting against a good recipient yields a bad reputation. They differ in how they treat behaviors of donors with bad recipients: Stern Judging (SJ) assigns bad to cooperating with a bad recipient and good to defecting against a bad recipient; Simple Standing (SS) treats any action toward a bad recipient as good; Shunning (SH) treats any action toward a bad recipient as bad; Scoring (SC) depends only on the action itself, that is, cooperation is good and defection is bad, so it is a first-order social norm. We assume $p$ represents the probability of gaining a good reputation by cooperating with a bad individual, and $q$ represents the probability of gaining a good reputation by defecting with a bad individual. Therefore, we can use $p$ and $q$ to parameterize these four norms, that is, to represent SJ, SS, SH and SC as $(0,1)$, $(1,1)$, $(0,0)$ and $(1,0)$ respectively. There is an assessment error with probability $u_{a}$ such that a good label can be flipped to bad and vice versa, where $0<u_{a}\ll 1/2$. The observer then broadcasts the assessment within the group so that group members share the same updated view of the focal individual. In what follows we set $u_{a}=u_{c}=0.02$.

\subsection{Coevolution of environment, strategy and norm}

The pair interactions and reputation updating operate on a fast time scale and alternate until reputations reach equilibrium. Under this standard assumption, if partner comparison does not depend on group identity, the strategy frequencies $f_{i}^{S}$ quickly converge across groups to a common value $f^{S}$. The subsequent evolutionary dynamics of strategies and norms follows the replicator form:
\begin{align}
\begin{split}
\dot f^{S}&=f^{S}(1-\tau)\bigl(\pi^{S}-\bar\pi\bigr),\qquad \\ \dot\nu_{i}&=\nu_{i}\tau\bigl(\pi_{i}-\bar\pi\bigr).
\end{split}
\end{align}
Here $\pi^{S}=\sum_{i} \nu_{i}\pi_{i}^{S}$, $\pi_{i}=\sum_{S} f^{S}\pi_{i}^{S}$, and $\bar\pi=\sum_{i} \nu_{i}\sum_{S} f^{S}\pi_{i}^{S}$ denote the average payoff of strategy $S$, the average payoff of group $i$, and the population average payoff. The parameter $\tau\in[0,1]$ allocates the slow-time weight between strategy replication and norm or group switching, so strategies update with weight $1-\tau$ and norms with weight $\tau$. In the extreme cases, $\tau=0$ yields pure strategy evolution and $\tau=1$ yields pure norm competition. When $0<\tau<1$, strategies and norms coevolve. Because we are mainly interested in how indirect reciprocity affects the level of cooperation under pure strategy evolution, and in how the gossip groups evolve under pure norm competition when the strategy is fixed (such as DISC), we therefore focus on the cases $\tau=0$ and $\tau=1$.

We further introduce environmental feedback to capture the bidirectional coupling between environmental resources and cooperation levels. Our basic assumption is that the evolution of the environmental state is determined by the net outcome of collective behaviors. Specifically, the environment improves only when the constructive synergy generated by cooperators outweighs the resource depletion caused by defectors, while it deteriorates when the destructive effects of defection dominate. Accordingly, the dynamics of the environmental state $n$ are governed by
\begin{align}
\dot n = \eta n(1-n) g(f_c)= \eta n(1-n)[\theta f_c-(1-f_c)].
\end{align}
Here $\eta$ is the relative rate of environmental change, and $\theta$ measures how sensitive the environment is to the population’s cooperation level. The current cooperation level $f_c$ can be expressed using strategies, group weights, and reputations:
\begin{align}
\begin{split}
f_c&=(1-u_c)\!\left(f^{\mathrm{ALLC}}+\sum_j \nu_j f_j^{\mathrm{DISC}} \sum_i \nu_i r_{j,i}\right) \\
&=(1-u_c)\!\left(f^{\mathrm{ALLC}}+f^{\mathrm{DISC}}\sum_j \nu_j r_{j,\cdot}\right).
\end{split}
\end{align}
The term $n(1-n)$ limits $n$ to between $0$ and $1$, so whether the environment improves or degrades is determined by the sign of $g(f_c)=\theta f_c-(1-f_c)$. If $g(f_c)>0$, the environment begins to recover; otherwise it is damaged.

Taken together, strategies, norms, and the environment form a closed coevolutionary system. Strategies and norms update through replicator dynamics, the environment evolves via the feedback equation, and all three are linked through payoffs and the cooperation level:
\begin{align}
\begin{cases} 
\dot f^{S}&=f^{S}(1-\tau)\bigl(\pi^{S}-\bar{\pi}\bigr),\\
\dot \nu_i&=\nu_i \tau\bigl(\pi_i-\bar{\pi}\bigr),\\
\dot n&=\eta\, n(1-n) g(f_c). 
\end{cases}
\end{align}

\section{Results}\label{sec3}
\subsection{Evolutionary dynamics of two strategies}

\begin{figure*}[htbp]
\centering
\includegraphics[width=1\textwidth]{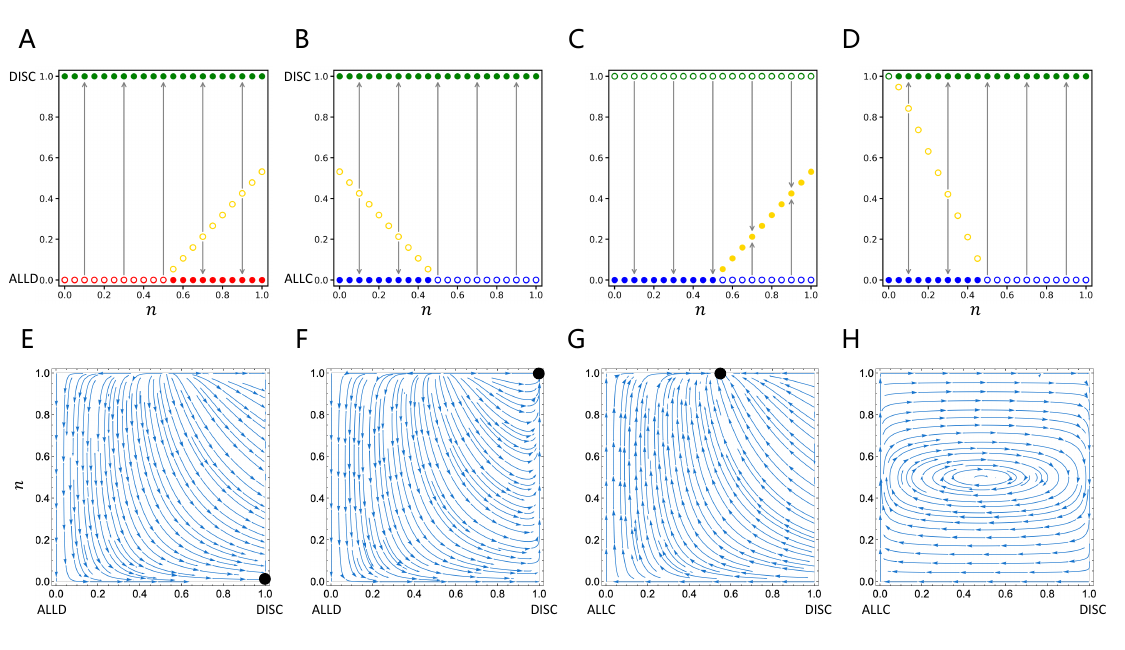}
\caption{\textbf{Two-strategy evolutionary dynamics in static and dynamic environments.} (A-D) Strategies are distinguished by color: green for DISC, red for ALLD, blue for ALLC, and yellow for coexistence. Solid and open circles denote stable and unstable equilibria, respectively.(A) Bistability between DISC and ALLD in a static environment ($n > 1/2$), where environmental scarcity facilitates DISC invasion. (B-D) DISC vs. ALLC in a static environment. SJ leads to bistability in poor environments (B), whereas SC supports stable coexistence in rich environments (C). Extreme parameters prevent DISC maintenance under SJ at low $n$ (D). (E-H) Impact of environmental feedback. The black solid dot indicates the stable equilibrium of the eco-evolutionary system. Feedback generally breaks the bistability in DISC vs. ALLD, promoting DISC dominance (E) and (F). In DISC vs. ALLC, feedback allows coexistence under SC (G) but triggers oscillations under SH (H). Parameters: $b=1.01$ in (D) and $b=2$ in all other panels; $p=0$, $q=1$ in (B) and (D); $p=1$, $q=0$ in (C); $q=0$, $\theta=2$ in (E); $q=0$, $\theta=5$ in (F); $p=1$, $q=0$, $\theta=2$ in (G); $p=0$, $q=0$, $\theta=0.5$ in (H).}\label{fig2}
\end{figure*}

Here we focus on the two-strategy dynamics with $\tau=0$ and $K=1$, which means all individuals are in a single well-mixed group and share the same social norm. Following our model, we first introduce a static environment $n$ that does not change over time and, under this assumption, examine the ability of DISC to invade and to resist invasion. In the competition between DISC and ALLD, solving $\dot f^{\mathrm{DISC}}=0$ yields an internal equilibrium $f_{\mathrm{ALLD}}^{\mathrm{DISC}}=\frac{c}{b}\frac{2n-1}{\epsilon-u_a}$. Here $\epsilon = (1-u_c)(1-u_a)+u_cu_a$ is the probability that an individual intends to cooperate with a good-reputation recipient and is judged to have a good reputation, which we denote by $P_{GC}$. Likewise, $P_{GD}=u_a,P_{BC}=p(\epsilon-u_a)+q(1-\epsilon-u_a)+u_a,P_{BD}=q(1-2u_a)+u_a$. A population of all defectors can resist invasion by discriminators if $n>\tfrac{1}{2}$, whereas a population of all discriminators can resist invasion by defectors if 
\begin{align}
\frac{b}{c}>\frac{2n-1}{\epsilon-u_a}
\end{align}
or $n<\frac{1+(b/c)(\epsilon-u_a)}{2}$. Since $u_c$ and $u_a$ are small, we have $\epsilon-u_a=(1-u_c)(1-2u_a)>0$. Naturally, the mutual resistance condition $\tfrac{1}{2}<n<\frac{1+(b/c)(\epsilon-u_a)}{2}$ guarantees a nonempty interval in which ALLD and DISC can resist each other’s invasion. Moreover, this interval coincides with the existence interval of $f_{\mathrm{ALLD}}^{\mathrm{DISC}}$, so this internal equilibrium is unstable. This indicates that DISC and ALLD display bistability in better environments (Fig.~\ref{fig2}A). In poorer environments ($n<\tfrac{1}{2}$), DISC can successfully invade ALLD because a poorer environment tends to favor cooperative choices, reduces ALLD’s vigilance against DISC, and thereby raises the level of group cooperation.

Similarly, for the competition between DISC and ALLC, the internal equilibrium is $f_{\mathrm{ALLC}}^{\mathrm{DISC}}=\frac{c}{b}\frac{2n-1}{P_{BC}-P_{BD}}$. This equilibrium does not exist under SH and SS, exists under SJ when $n<\tfrac{1}{2}$, and exists under SC when $n>\tfrac{1}{2}$. A population of all cooperators can resist invasion by discriminators if $n<\tfrac{1}{2}$, whereas a population of all discriminators can resist invasion by cooperators if $n>\frac{1+\frac{c}{b}(P_{BC}-P_{BD})}{2}$. Equivalently, in terms of $b/c$,
\begin{align}
\begin{cases}
n>\dfrac{1}{2}, & \text{under SH and SS},\\[5pt]
\dfrac{b}{c}>\dfrac{2n-1}{u_a-\epsilon,}, & \text{under SJ},\\[5pt]
\dfrac{b}{c}<\dfrac{2n-1}{\epsilon-u_a,}, & \text{under SC}.
\end{cases}
\end{align}
However, $f_{\mathrm{ALLC}}^{\mathrm{DISC}}$ is unstable under SJ and stable under SC. Hence, SJ yields bistability when $n<\tfrac{1}{2}$ (Fig.~\ref{fig2}B), SC yields a stable interior equilibrium when $n>\tfrac{1}{2}$ (Fig.~\ref{fig2}C), and SH/SS yield only boundary monostability. In an extreme case with very small $b$ (such as $b=1.01$), DISC cannot resist invasion by ALLC under SJ when $n$ approaches $0$ (Fig.~\ref{fig2}D), while under SC it can resist when $n$ approaches $1$.

In a static environment $n$, the two kinds of contests display different sensitivities. We can regard $n$ as a uniform weighting applied to the existing evaluation rule. The rule stays fixed, and $n$ only scales how much pre-existing differences are amplified. For ALLD–DISC, both strategies take the same attitude toward B (bad-reputation recipients). Normative differences act mainly along this dimension, yet they cancel in a relative comparison because the two strategies move in the same direction. The environment therefore only magnifies or attenuates the baseline gap and the outcome is insensitive to the norm (see Fig. S1 in Supplementary Information). For ALLC–DISC, the strategies share the same attitude toward G (good-reputation recipients) but take opposite attitudes toward B. The norms assign different evaluations to these opposite attitudes, and the environment’s uniform weighting of that evaluation directly shifts which strategy has the advantage, leading to a pronounced dependence on the norm (see Fig. S2 in Supplementary Information).

When the environmental state $n$ is affected by the group cooperation level $f_c$ and varies over time, the coevolutionary outcomes of the two-strategy game differ markedly from those in a static environment where $n$ is fixed. Here, we focus on the competitive dynamics between ALLD and DISC and between ALLC and DISC. Numerical simulations show that, compared with the typical bistable structure observed in static environments (for instance, the bistability between full DISC and full ALLD), introducing environmental feedback greatly amplifies the evolutionary advantage of DISC. Over a wide range of parameter values, the system no longer remains in a bistable regime where DISC coexists or competes with the other strategy, but instead tends to converge uniquely to DISC (Fig.~\ref{fig2}E and Fig.~\ref{fig2}F).

In the competition between ALLD and DISC, an interesting phenomenon emerges: for a given value of $q$, the evolutionary dynamics are almost identical (see Fig. S4 in Supplementary Information). In other words, under dynamic environments, SJ and SS on the one hand, and SC and SH on the other, produce pairwise identical evolutionary outcomes. Intuitively, this indicates that what determines the direction of evolution is how defections against badly reputed individuals are evaluated, rather than how cooperation with such individuals is evaluated. This can be understood more clearly from a mathematical perspective. In the ALLD–DISC competition, there are no unconditional cooperators (ALLC) in the population. Once an individual is labeled as having bad reputation, both ALLD and DISC always defect against them, and events of “cooperating with a bad individual” are essentially absent. The corresponding path probability $P_{BC}=p\left(\epsilon-u_a\right)+q\left(1-\epsilon-u_a\right)+u_a$ therefore does not effectively participate in the evolutionary process. The parameter $p$ appears only in $P_{BC}$, while in the other three reputation-update probabilities only $q$ enters. This implies that, in the ALLD–DISC competition, $q$ is the key parameter that determines whether DISC can maintain a good reputation in the long run, thereby controlling the ranking of average payoffs. Similarly, the relative response speed $\eta$ mainly affects the convergence rate rather than the final equilibrium position, whereas an increase in $\theta$ tends to drive the environment from a degraded state to a more abundant one (Fig.~\ref{fig2}E and Fig.~\ref{fig2}F). It is worth noting that although DISC is globally attracting in most cases under ALLD–DISC competition, it is not the only possible long-term behavior. Under extreme parameter conditions (see Fig. S4 in Supplementary Information, when the benefit factor $b$ is very close to $1$, such as $b=1.01$), the system can exhibit a heteroclinic cycle connecting multiple boundary states. In this regime, when DISC temporarily gains an advantage, the rising cooperation level improves the environment; however, under nearly neutral selection, ALLD can again invade by exploiting the improved environment, gradually dragging it back to a poor state. As the environment becomes excessively harsh, the relative payoff of DISC improves once more, leading to its resurgence. Consequently, the system cycles among several quasi-steady states, manifesting as persistent oscillations in both strategies and environmental quality.

In the competition between ALLC and DISC, the dynamics become more intricate. Because both the reputation update and the payoff structure now depend simultaneously on $p$ and $q$, differences among social norms are fully amplified. Overall, DISC remains the unique attractor in most regions of parameter space (see Fig. S5 in Supplementary Information, under SJ and SS, and under SH with big $\theta$), but under the SC norm long-term coexistence between ALLC and DISC becomes possible (Fig.~\ref{fig2}G). Intuitively, SC is relatively permissive in how it evaluates actions toward badly reputed individuals: on the one hand, defection against a bad individual is not always rewarded; on the other hand, cooperation with a bad individual can still receive a positive evaluation. This compromise rule provides both ALLC and DISC with viable pathways to maintain good reputation, so that neither strategy can completely dominate the other in terms of long-run payoffs, and an internal coexistence equilibrium emerges in the dynamics. In addition, for small values of $\theta$ we observe pronounced oscillations under the SH norm (Fig.~\ref{fig2}H). SH evaluates behaviors toward badly reputed individuals in a generally harsher manner, so that once the environment slightly deteriorates, both strategies suffer a combined penalty in reputation and payoff; together with the weaker positive environmental feedback (small $\theta$), this can generate persistent cycling. Because the presence of ALLC substantially raises the overall level of cooperation, even for small $\theta$ the environment tends to oscillate within a not-too-degraded range or settle at a relatively abundant state, and its average quality is markedly higher than in the case with ALLD–DISC competition alone.

\subsection{Evolutionary dynamics of three strategies}

\begin{figure*}[t]
\centering
\includegraphics[width=1\textwidth]{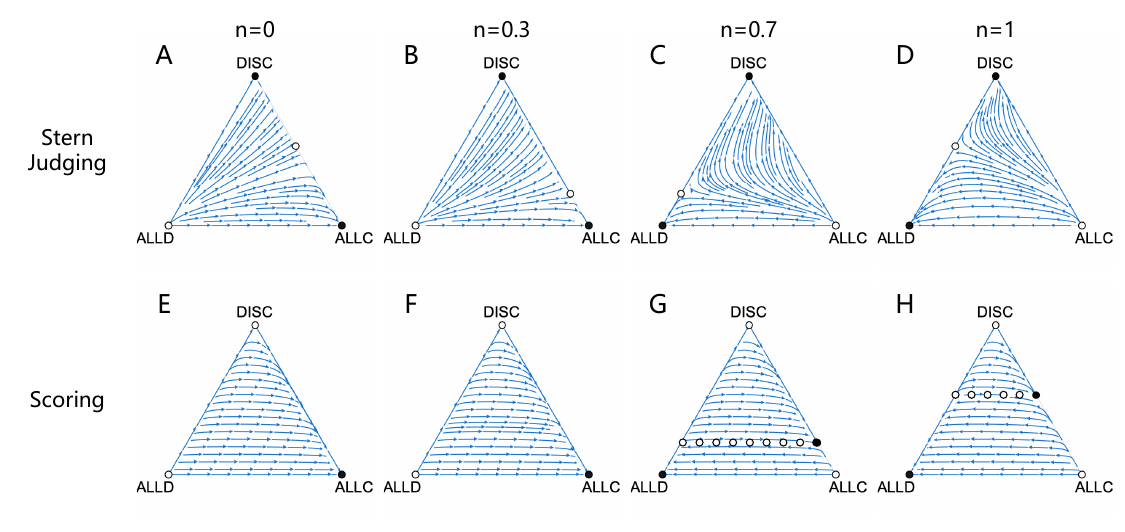}
\caption{\textbf{Evolutionary dynamics of three strategies in static environments.} The simplexes illustrate evolutionary trajectories under two social norms (rows) and varying static environments ($n$, columns). Solid circles denote stable equilibria, and open circles denote unstable equilibria. (A–D) Under Stern Judging, the system transitions from a bistable configuration of ALLC and DISC in poor environments ($n < 1/2$) to a bistability between ALLD and DISC in rich environments ($n > 1/2$). (E–H) Under Scoring, ALLC is globally stable in poor environments. In rich environments, the system exhibits a unique bistability between a pure ALLD equilibrium and a stable boundary coexistence of DISC and ALLC. Parameter: b=2.}\label{fig3}
\end{figure*}

We now turn to the evolutionary dynamics when all three strategies are present simultaneously. We retain the setting from the previous section, taking $\tau=0$ and $K=1$, so that all individuals belong to a single group and share the same social norm. In this framework, only the strategies change among the three options, while the norm itself remains fixed. Although the evolution of three-strategy indirect reciprocity has been studied relatively systematically in the literature, comparing the dynamics under static environments and under environmental feedback helps us understand the stability and diversity of indirect reciprocity in more realistic scenarios.

We first examine the evolutionary dynamics in a static environment. In this case, during repeated interactions, individual reputations and strategies can change over time, but the environmental state $n$ and the resulting payoff structure remain fixed. When $n=1$, the model reduces to the classical three-strategy donation game. Consistent with previous findings, under the SJ, SS, and SH norms, DISC and ALLD form a bistable configuration, while ALLC typically corresponds to an unstable equilibrium and cannot persist as a standalone strategy in the long run (Fig.~\ref{fig3}D, and Fig. S6 in Supplementary Information).

Changes in the static environment $n$ reshape the equilibrium structure of the three-strategy system. When the environment is relatively abundant (e.g., $n > 1/2$), under SJ, SS, and SH, the bistable structure between DISC and ALLD still exists (Fig.~\ref{fig3}C, and Fig. S6 in Supplementary Information). Within one basin of attraction, DISC can maintain an internal cooperative equilibrium by relying on higher cooperation levels; while in the other basin of attraction, the high payoffs from defection are sufficient to support a pure ALLD equilibrium. Under the SC norm, the three strategies exhibit a different bistability: a pure ALLD equilibrium and a boundary coexistence equilibrium composed of DISC and ALLC (Fig.~\ref{fig3}G and Fig.~\ref{fig3}H). Furthermore, multiple unstable equilibrium points exist within the simplex, which constitute the boundary separating different basins of attraction. This special boundary coexistence originates from the first-order assessment mechanism of the SC norm: SC updates reputation solely based on the action itself, making ALLC and DISC phenotypically indistinguishable in a highly cooperative environment. Therefore, both can jointly resist the invasion of defectors but cannot exclude each other through mutual competition internally.

When the static environment is poor, the steady-state structure of the three-strategy system changes markedly. On the one hand, under SJ, ALLC and DISC form a bistable configuration (Fig.~\ref{fig3}A and Fig.~\ref{fig3}B). On the other hand, under SS, SC and SH, ALLC becomes the only stable equilibrium (Fig.~\ref{fig3}E-F, and Fig. S6 in Supplementary Information). In other words, in a degraded environment, pure cooperators can not only successfully invade populations dominated by other strategies, but also maintain resistance to invasion in the long run, while ALLD no longer constitutes a viable equilibrium under any of the four norms. The reason is that, in a poor environment, the additional payoff that defection relies on is greatly reduced, whereas ALLD continues to suffer from reputational sanctions, making it difficult to sustain a positive net growth rate. By contrast, ALLC always receives a positive evaluation when interacting with well-reputed individuals, and under resource scarcity, mutual cooperation becomes one of the few interaction patterns that can still yield relatively high payoffs. The contribution of cooperation among ALLC players is therefore strongly amplified, so that the pure ALLC state is strongly self-sustaining under SS, SC, and SH.

Moreover, we find that the stability of DISC is jointly constrained by the type of norm and the environmental state. Under SJ, DISC can exist as a pure-strategy equilibrium (Fig.~\ref{fig3}A-D). When the environment is favorable and defection yields high payoffs, pure ALLD remains self-consistent, while DISC accumulates reputation through cooperating with good recipients and sanctioning bad ones, and maintains a higher average payoff, leading to a bistable configuration between DISC and ALLD. When the environment is poor and cooperation becomes relatively more profitable, the pure ALLC equilibrium becomes sustainable. However, SJ still provides explicit reputational rewards for DISC, so that, given a sufficiently large initial fraction, DISC can also persist as a pure-strategy equilibrium, and the system then exhibits bistability between DISC and ALLC. Under SS and SH, the incentives for punishing bad recipients are weaker than under SJ, so DISC has enough payoff and reputational advantage to form a local bistable configuration with ALLD only when the environment is favorable (Fig. S6 in Supplementary Information). Once the environment becomes degraded, the payoff advantage of cooperative strategies is amplified, and the always-cooperating ALLC becomes dominant under these norms, while DISC turns unstable under SS, SC, and SH (Fig.~\ref{fig3}E-F, and Fig. S6 in Supplementary Information). The case of SC is particularly distinctive: reputation depends only on whether the donor cooperates and does not distinguish the recipient, so DISC has no additional advantage over ALLC, and it is also difficult for DISC to fully eliminate ALLD in a rich environment. As a result, DISC cannot support a pure-strategy equilibrium under SC (Fig.~\ref{fig3}G and Fig.~\ref{fig3}H).

\begin{figure*}[t]
\centering
\includegraphics[width=1\textwidth]{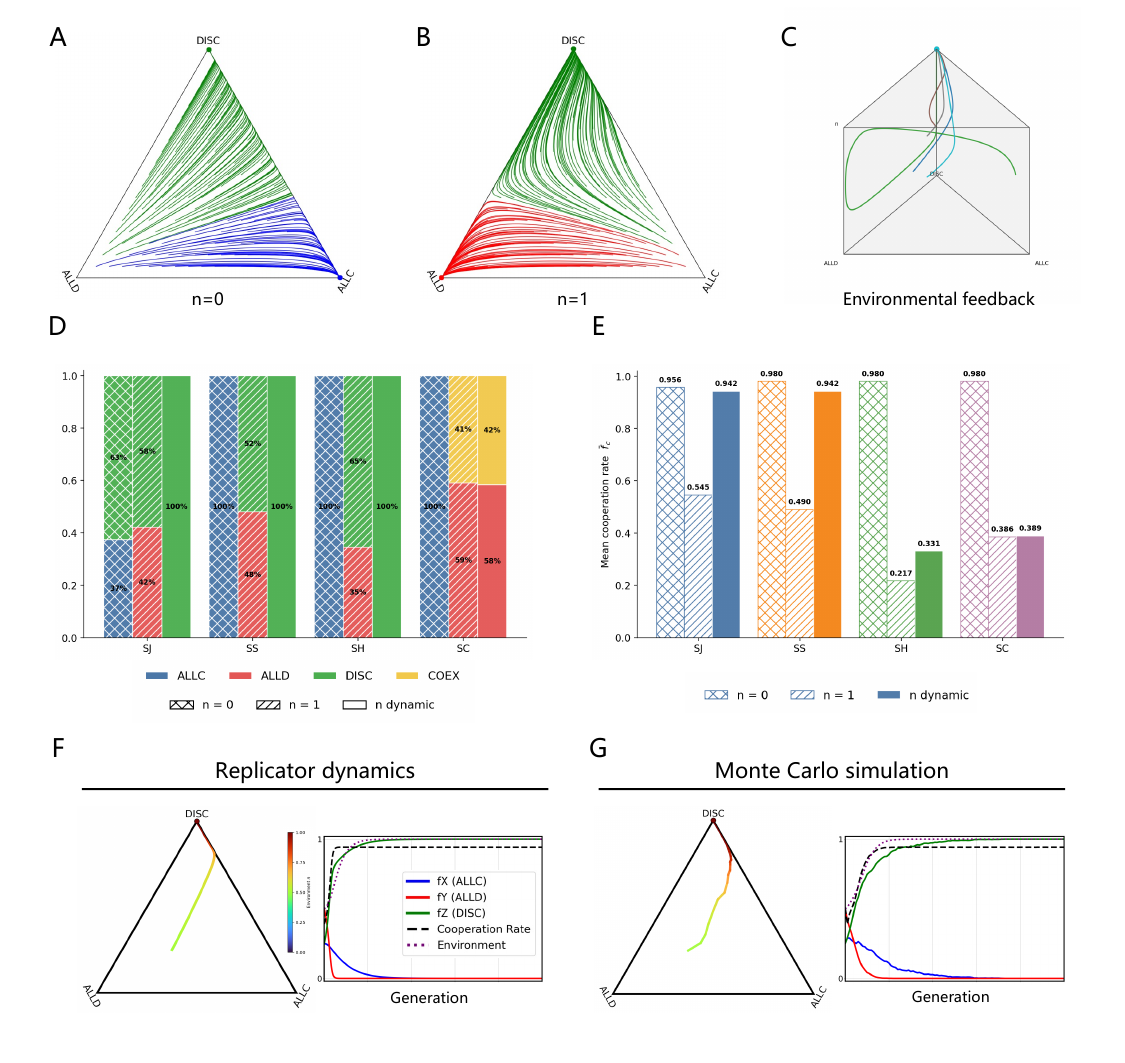}
\caption{\textbf{Impact of environmental feedback on three-strategy evolutionary dynamics.} (A–C) Evolutionary trajectories on the simplex under Stern Judging in static poor ($n=0$, A), static rich ($n=1$, B), and dynamic environments (C), illustrating how feedback alters the basins of attraction. (D) Comparison of steady-state strategy frequencies across four norms under static ($n=0, n=1$) and dynamic conditions. (E) Corresponding average cooperation levels for the scenarios in (D). (F–G) Temporal evolution of strategy frequencies and environmental state ($n$). The numerical solutions of the replicator dynamics (F) show excellent agreement with agent-based Monte Carlo simulations (G). Parameters: $b=3, \theta=3, \epsilon=0.1$.}\label{fig4}
\end{figure*}

Furthermore, the environment functions not as a static backdrop but as a dynamic entity, continuously modulated by the level of cooperation within the population, which in turn alters the relative payoffs of different strategies. Consequently, it is essential to integrate the environmental variable into the dynamical framework, treating strategy frequencies and the environmental state as a coupled co-evolving system. The subsequent analysis focuses on the dynamics under the SJ norm, widely recognized as the strongest among the four social norms considered (results of other three norms are shown in Fig. S7 in Supplementary Information).

First, in the static environment, the system exhibits bistability between DISC (green) and ALLC (blue) at $n=0$ (Fig.~\ref{fig4}A), whereas at $n=1$, the bistability shifts to DISC and ALLD (red) for $b=3$ (Fig.~\ref{fig4}B). This pattern is qualitatively consistent with the steady-state structure previously reported for the static case with $b=2$ (Fig.~\ref{fig3}A and Fig.~\ref{fig3}D). Upon introducing environmental feedback under the SJ norm, the dynamics change fundamentally: the system converges to a globally stable pure DISC equilibrium (at $\theta=3$) regardless of the initial state (Fig.~\ref{fig4}C). Monte Carlo simulations in finite populations corroborate these replicator dynamics results (Fig.~\ref{fig4}F and Fig.~\ref{fig4}G), confirming that environmental feedback significantly expands the DISC basin of attraction, effectively eliminating both ALLC and ALLD to establish DISC as the unique long-term equilibrium.

Similarly, under the SS and SH norms, the introduction of environmental feedback establishes DISC as the unique stable equilibrium (see Fig. S7 in Supplementary Information). This contrasts markedly with the static extremes, where a pure ALLC equilibrium prevails in poor environments ($n=0$) and DISC–ALLD bistability emerges in rich environments ($n=1$), indicating that eco-evolution significantly reinforces the dominance of DISC. Conversely, the dynamics under the SC norm present a distinct scenario. Compared with the favorable static environment ($n>1/2$), after incorporating environmental feedback, the type of long-term behavior does not change fundamentally. ALLD can still exist as an attractor, while the other type of attractor consists of the coexistence of DISC and ALLC. Notably, convergence is significantly retarded: for a subset of initial conditions, trajectories exhibit prolonged quasi-periodic fluctuations before ultimately converging to the ALLD attractor. Thus, compared to adverse static environments ($n<1/2$) characterized solely by pure ALLC, the dynamic environment fosters a richer landscape of long-term evolutionary outcomes under the SC norm (see Fig. S7 in Supplementary Information).

We quantify the relative dominance of strategies by measuring the size of their basins of attraction for each norm-environment combination (Fig.~\ref{fig4}D). Accordingly, the average cooperation level is defined as $\bar{f_c}=\sum_{i}{p_if_c^{(i)}}$, where $p_i$ represents the basin size of the $i$-th attractor and $f_c^{(i)}$ denotes its associated cooperation level. In the resource-poor static environment ($n=0$), selection pressure overwhelmingly favors ALLC, resulting in maximal $\bar{f_c}$ across all four norms. Conversely, in the resource-rich static setting ($n=1$), ALLD exploits its payoff advantage while DISC leverages reputational benefits; their combined dominance suppresses pure cooperators, maintaining low overall cooperation. 

Since the payoff structure at $n=1$ degenerates into the standard donation game, we establish this state as the canonical baseline for comparison. Introduction of environmental feedback elevates $\bar{f_c}$ above this baseline for all norms, with the most pronounced enhancements observed under SJ, SS, and SH (Fig.~\ref{fig4}E). Although cooperation in the dynamic case falls below the peak levels seen in the static $n=0$ scenario, which represents the theoretical upper bound driven by survival necessity, the decline under SJ and SS is marginal, preserving the system in a high-cooperation regime. In contrast, SH and SC suffer more substantial reductions. This deficiency arises because their assessment rules regarding interactions with ill-reputed individuals are either too coarse or excessively severe, rendering them less effective at sustaining robust, fine-grained conditional cooperation. In summary, environmental feedback effectively raises cooperation levels relative to the donation-game baseline and promotes cooperative robustness across a broad range of conditions.

In the results described above, the environment ultimately evolves toward saturation. This occurs because a large environmental sensitivity $\theta$ amplifies the positive feedback from average cooperation. Provided cooperation remains above a critical threshold, the recovery term dominates, driving $n$ to its upper bound. However, as $\theta$ diminishes, this driving force attenuates, potentially altering both the environmental steady state and the associated strategic balance (see Fig. S8 in Supplementary Information). Under the SJ and SS norms, which inherently sustain high cooperation levels, reducing $\theta$ neither significantly lowers the long-term environmental level nor qualitatively modifies the equilibrium structure. In contrast, under the SH norm—characterized by lower baseline cooperation—a decrease in $\theta$ renders the environmental feedback insufficient to counterbalance the degradation induced by defection. Consequently, the system exhibits persistent oscillations in both environmental state and strategy composition over prolonged time scales. Notably, ALLD remains excluded consistent with the strict reputational sanctions against pure defectors; thus, these oscillations are confined primarily to the ALLC–DISC subspace. The dynamics under the SC norm are distinct. At relatively high $\theta$, the system displays long-lasting quasi-periodic oscillations that eventually settle into a boundary equilibrium. As $\theta$ decreases further, these oscillating trajectories contract toward the interior and stabilize as multiple interior equilibria. This implies that all three strategies can coexist at various ratios, with the environment converging to distinct intermediate steady states rather than being forced to the saturation limit.

\subsection{Evolutionary dynamics of social norms}

\begin{figure*}[htbp]
\centering
\includegraphics[width=1\textwidth]{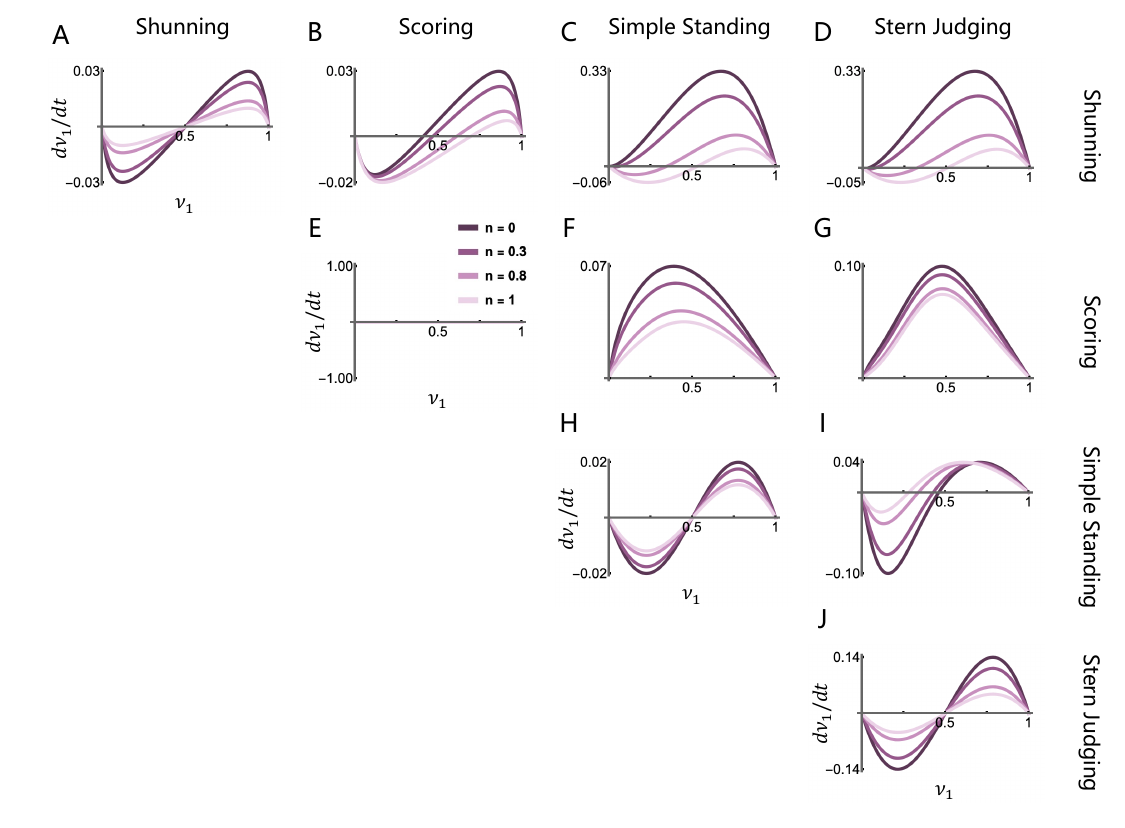}
\caption{\textbf{Norm competition in static environments.} The panels plot the time derivative of Group 1’s fraction, $\dot{\nu}_1$ (vertical axis), against its current fraction, $\nu_1$ (horizontal axis), under different static environmental conditions $n$. The dynamics typically exhibit bistability, where the intersection with the horizontal axis, denoted as $\nu_1^*$, represents the unstable threshold separating the basins of attraction for the two competing norms. The relative competitiveness of a norm is visually determined by the position of this threshold: if $\nu_1^* < 1/2$, the norm adopted by Group 1 possesses a larger basin of attraction and can take over the population even from a minority size. Parameter: $b=2$.}\label{fig5}
\end{figure*}

In the preceding analysis, we focused on the coevolution of strategies and the environment under a fixed social norm. We now shift the perspective to the evolution of norms. We fix $\tau=1$ and assume that all individuals in the population adopt the DISC strategy, so individual strategies no longer evolve and the objects of evolution become the groups that carry different social norms, with reputation serving as the sole criterion for behavioral evaluation. Specifically, the model contains $K=2$ assessment groups, each of which applies a given social norm internally; the two groups may follow the same norm or two different norms. In this setting, norm evolution is described by the expansion and contraction of the group sizes associated with each norm, that is, by the relative competition between groups driven by differences in reputation and payoff.

In line with previous studies, we find that the evolution of group sizes with a static environment under norm competition typically exhibits a bistable structure. The boundary states $\nu_1=0$ and $\nu_1=1$ are both stable equilibria, and there exists an intermediate unstable threshold $\nu_1^{*}$. If the initial group size $\nu_1$ is larger than this threshold, the system evolves toward $\nu_1=1$; otherwise, it evolves toward $\nu_1=0$. When $\nu_1^{*}<1/2$, the norm used by Group 1 is more likely to take over the entire population, even if Group 1 accounts for only half or less of the population initially. This motivates us to examine the sign of $\dot{\nu}_1$ at the symmetric state $\nu_1=1/2$. If $\left.\dot{\nu}_1\right|_{\nu_1=1/2}>0$, then $\nu_1^{*}<1/2$ necessarily holds, implying that Group~1 enjoys a population-level advantage in norm competition.

By setting $\nu_1=\nu_2=1/2$, we obtain that $\left.\dot{\nu}_1\right|_{\nu_1=1/2}>0$ if and only if
\begin{multline}
\Big[\big(b-(2n-1)c\big)\big(g_{1,1}-g_{2,2}\big) \\
+\big(b+(2n-1)c\big)\big(g_{1,2}-g_{2,1}\big)\Big]
\Big|_{\nu_1=\frac{1}{2}} > 0 .
\end{multline}
The first term, $\big(b-(2n-1)c\big)\big(g_{1,1}-g_{2,2}\big)$, represents the payoff difference between the two norms in within-group interactions, whereas the second term, $\big(b+(2n-1)c\big)\big(g_{1,2}-g_{2,1}\big)$, captures their payoff difference in between-group interactions. The environmental state $n$ modulates these contributions through the coefficients $b\pm(2n-1)c$: when the environment is favorable (large $n$), the weight on between-group differences is amplified, and when the environment is poor (small $n$), the weight on within-group differences becomes more pronounced. Equivalently, a favorable environment tends to favor norms that perform better in out-group encounters, while a poor environment tends to favor norms that are more effective at maintaining in-group cooperation and reputation. If the resulting weighted sum is positive, then starting from the symmetric initial condition, the norm adopted by Group~1 will gradually expand through group-level competition and eventually dominate the entire population.

For norm dynamics, the evolving entities are the two groups that carry different social norms, with sizes $\nu_1$ and $\nu_2=1-\nu_1$. At the initial time, if we assume that the two groups are of equal size, $\nu_1=\nu_2=1/2$, then the position of the unstable threshold $\nu_1^{*}$ characterizes the relative strength of the two norms. When $\nu_1^{*}<1/2$ (equivalently, when the above inequality condition holds), the norm adopted by Group~1 has a competitive advantage at the group level. Intuitively, even if Group~1 starts from a smaller size, it can still induce the other group to switch to its own norm. When the two groups adopt exactly the same norm, we have $\nu_1^{*}=1/2$, in which case the norms are equivalent in terms of fitness and the group with the larger initial size will dominate the population in the long run (Fig.~\ref{fig5}A, Fig.~\ref{fig5}E, Fig.~\ref{fig5}H and Fig.~\ref{fig5}J).

When the two groups adopt different social norms, the threshold condition above can be used to characterize the relative strength of the competing norms. In our model, SJ remains the strongest among the four norms and typically outcompetes the other three ($b=2$, (Fig.~\ref{fig5}D, Fig.~\ref{fig5}G and Fig.~\ref{fig5}I)). Moreover, the environmental state systematically modulates this advantage structure. When the environment is relatively poor, the advantage of SJ over SH is substantially amplified (Fig.~\ref{fig5}D). Under SJ, donors distinguish between different types of recipients and impose sanctions only on genuinely inappropriate actions, which effectively suppresses opportunistic behavior while avoiding unjust penalties on justified punishment or necessary sanctioning. It allows the cooperative strategy to fully exploit its relative payoff advantage in such environments. In contrast, SH applies a uniform exclusion rule to all individuals with bad reputation, which leads to a severe loss of potential partners and a marked reduction in the average group payoff. By comparison, SJ and SS treat badly reputed individuals in a more similar manner. As a result, their fitness difference becomes smaller in poor environments, and the advantage of SJ over SS decreases as the environment deteriorates (Fig.~\ref{fig5}I).

\begin{figure*}[htbp]
\centering
\includegraphics[width=1\textwidth]{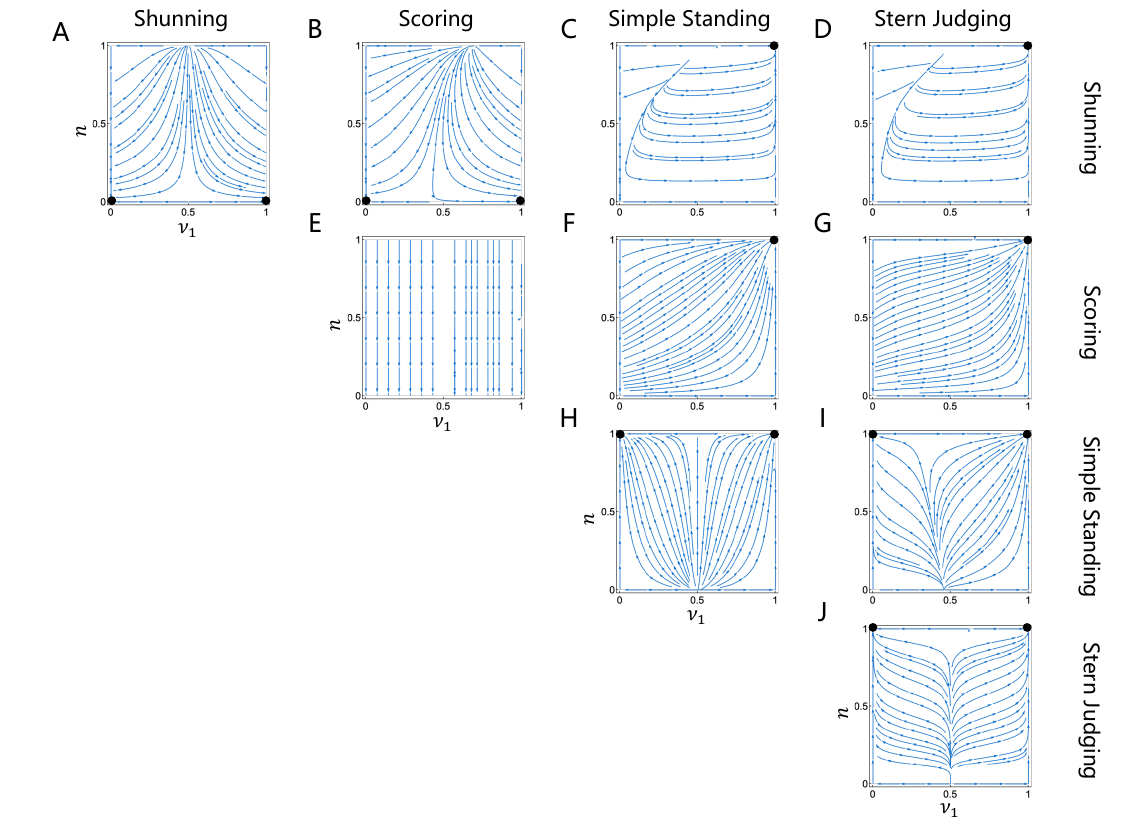}
\caption{\textbf{Phase diagrams of pairwise competition between social norms in a dynamic environment.} 
The columns and rows correspond to the social norms adopted by Group 1 and Group 2, respectively. In each panel, the horizontal axis represents the population fraction of Group 1 ($\nu_1$), and the vertical axis represents the environmental resource level ($n$). The stream plots illustrate the direction of selection driven by the coupled dynamics of strategies and environmental feedback. Black dots denote stable equilibria (attractors) where the system settles in the long run. Parameters: $b=2, \theta=2, \epsilon=0.1$.}\label{fig6}
\end{figure*}

In contrast, the performance of SC is markedly different. The conventional view is that SC is often the weakest of the four norms, because it evaluates actions solely based on whether the donor cooperates and does not take the recipient’s reputation into account. As a result, it cannot specifically punish helping bad recipients or reward sanctioning them, and in most situations it is dominated by more discriminating norms. In our model with a static environment, however, we identify an exception: when the environment is poor, SC can outperform SH (Fig.~\ref{fig5}B). Intuitively, SH almost completely excludes badly reputed individuals from future interactions, which, in a degraded environment, drastically shrinks the available cooperation network and causes the group to forgo profitable cooperative opportunities. In contrast, the more permissive evaluation under SC preserves a larger set of potential reciprocal partners and thereby yields a higher average in-group payoff under poor environmental conditions. Nonetheless, SC never surpasses SJ or SS in any environmental state (Fig.~\ref{fig5}F and Fig.~\ref{fig5}G). Its lack of a precise and consistent punishment scheme for defectors leads to systematically lower long-term performance compared with these two norms.

In the competition with SH, SJ and SS retain a clear overall advantage (Fig.~\ref{fig5}C and Fig.~\ref{fig5}D). As the environment deteriorates and cooperation becomes relatively more profitable, this advantage is further amplified, and differences in their ability to sustain cooperation and repair reputation within groups become more pronounced. Even when the initial fraction of individuals adopting SH is high, the long-term dynamics still tend to be dominated by SJ or SS. In contrast, when the environment is favorable and defection is more attractive, SH can rival these two norms and may even gain a slight advantage.

When the environment changes from a static parameter to a dynamic variable, the coevolutionary pattern of norms and environment is systematically altered. First, when the two groups follow the same social norm, the outcome remains identical to the static case: the group that is initially larger eventually occupies the entire population (Fig.~\ref{fig6}A, Fig.~\ref{fig6}E, Fig.~\ref{fig6}H and Fig.~\ref{fig6}J). Although SJ is already the most advantageous norm under a static environment, environmental feedback further amplifies its dominance. In pairwise competition between SJ and SH or between SJ and SC, the phase diagram is fully dominated by SJ (Fig.~\ref{fig6}D and Fig.~\ref{fig6}G). In the competition between SJ and SS, a bistable structure persists, but the basin of attraction of SJ is clearly larger than that of SS (Fig.~\ref{fig6}I). At the same time, environmental feedback also improves the performance of SS relative to SC and SH, so that SS likewise exhibits a single attracting equilibrium in its competitions with SC and with SH (Fig.~\ref{fig6}C and Fig.~\ref{fig6}F). Finally, in the competition between SC and SH, the system typically displays a bistable pattern, and the basins of attraction associated with the two norms are of comparable size (Fig.~\ref{fig6}B).

In our coevolutionary framework, the long-term environmental quality is fundamentally determined by the capacity of each social norm to sustain cooperation. Numerical results show that SC and SH tend to keep the environment at a relatively low level (Fig.~\ref{fig6}A, Fig.~\ref{fig6}B and Fig.~\ref{fig6}E). while SJ and SS are more likely to maintain $n$ in a relatively favorable range (Fig.~\ref{fig6}C, Fig.~\ref{fig6}D and Fig.~\ref{fig6}F-J). The reason is that SJ and SS can support higher and more stable cooperation levels in poor environments, which provides a persistent positive feedback on environmental recovery. In addition, once environmental feedback is introduced, SJ and SS gain a fitness advantage over SC and SH in norm competition, leading to a single attracting equilibrium or a much larger basin of attraction As a consequence, groups following SC or SH are gradually eliminated in the long run, and their detrimental impact on the cooperation network and on the environment is progressively reduced (Fig.~\ref{fig6}C, Fig.~\ref{fig6}D, Fig.~\ref{fig6}F and Fig.~\ref{fig6}G).

Finally, we examine how the environmental feedback rate $\eta$ and the environmental sensitivity $\theta$ affect the coevolutionary outcomes of norms and environment. For a given pair of competing norms, varying $\eta$ does not change the long-term equilibria of the system, while it only rescales the time needed to approach these equilibria and slightly modifies the transient trajectories. In contrast, $\theta$ has a substantial impact on the steady-state structure, since it directly measures the strength of the positive effect of cooperation on environmental recovery and thus shifts the environmental equilibrium under different norm combinations. Specifically, increasing $\theta$ triggers a bifurcation from global SJ dominance to a bistable configuration in the SJ–SH competition (Fig. S9 in Supplementary Information), whereas it drives a critical transition from environmental poverty to abundance within the persistent bistable structure of the SC–SH competition (Fig. S10 in Supplementary Information).

\section{Conclusion}
Indirect reciprocity has long been regarded as an evolutionary cornerstone of large-scale human cooperation~\cite{nowak2006five}. However, classical theories predominantly situate games within a static environmental backdrop, overlooking the profound bidirectional feedback between behavioral strategies and public resources. By constructing a co-evolutionary framework coupling strategies, norms, and the environment, we demonstrate that environmental feedback constitutes not merely a parametric perturbation but introduces a fundamental systemic selection pressure. Such endogenous dynamics transform cooperation from a contingent outcome dependent on initial conditions into an evolutionary necessity capable of actively reshaping the environment to lock in its own advantage. This insight resonates with and extends the emerging discourse on eco-evolutionary game theory~\cite{wang2020eco}.

Specifically, in static, resource-rich models, the persistence of cooperation is often constrained by bistable dynamics, characterized by a significant dependence on initial conditions. In sharp contrast, the SJ norm~\cite{pacheco2006stern} demonstrates exceptional evolutionary robustness under dynamic feedback. Once environmental sensitivity exceeds a critical threshold, a potent positive feedback loop emerges between strategies and the environment: cooperative behavior restores the environment, and while a superior environment heightens the temptation to defect, it simultaneously amplifies the relative fitness of the discriminatory punishment mechanism inherent to SJ. This mechanism effectively eliminates the uncertainty found in static models and drastically expands the basin of attraction for the reputation-based discriminator strategy (DISC), rendering it the globally unique evolutionary endpoint across a broad parameter space and thereby exerting a locking effect analogous to ``environmental engineering".

Despite the general dominance of SJ, our static analysis reveals a significant context-dependency in norm performance, uncovering a counter-intuitive ``paradox of tolerance." In extremely resource-depleted static environments, the SC norm, often regarded as lacking discriminative power, surprisingly outperforms the stricter SH norm. This phenomenon arises because the ``zero-tolerance" policy of SH leads to a total collapse of cooperative networks under adverse conditions, incurring prohibitive opportunity costs. In contrast, the imprecise evaluation of SC, while lacking precision, preserves precious interaction opportunities during times of scarcity. This finding challenges the conventional wisdom that higher-order discrimination is universally superior to simple scoring, suggesting an evolutionary trade-off regarding optimal moral standards between stages of survival crisis and prosperity~\cite{ohtsuki2004should}.

Nonetheless, once viewed through a co-evolutionary lens, this transient survival advantage of the SC norm effectively vanishes. Our results establish a definitive dynamic hierarchy: SJ and SS consistently outperform SC and SH. The failure of the latter two stems not merely from the lower average cooperation levels they sustain, but more fundamentally from their inherent dynamic instability. Lacking the capacity to precisely sanction defection without alienating cooperators, SC and SH drive the system into long-lasting quasi-periodic oscillations rather than stable convergence, failing to maintain the positive environmental feedback required for long-term prosperity. Consequently, they are eventually displaced during norm competition by SJ, which possesses a superior environmental carrying capacity. This underscores that only norms with high discriminative power can underpin the long-term sustainability of eco-social systems.

While premised on assumptions of well-mixed populations and timescale separation that simplify certain complexities of real-world systems, our model offers a novel ecological perspective on the emergence of cooperation. Future inquiries could further explore the impact of spatial structure and multi-polar gossip groups on environmental feedback by examining how localized interaction networks induce nonlinear norm propagation effects~\cite{henrich2001search, li2020evolution, wang2024evolutionary}. Moreover, relaxing the assumption of timescale separation between strategy evolution and norm transmission to explore fully coupled dynamics on synchronous timescales may reveal richer systemic behaviors~\cite{kessinger2023evolution}. Furthermore, moving beyond deterministic frameworks to construct non-linear feedback modes incorporating ecological tipping points or stochastic perturbations would help uncover the resilience of social systems under extreme environmental conditions~\cite{hauert2019asymmetric, otto2020social}. Ultimately, endogenizing the environmental dimension into game-theoretic dynamics not only expands the boundaries of evolutionary game theory but also offers theoretical insights for addressing the increasingly critical tragedy of the commons.

%\noindent
%The input format for the above table is as follows:

%%=============================================%%
%% For presentation purpose, we have included  %%
%% \bigskip command. Please ignore this.       %%
%%=============================================%%

%%=============================================%%
%% For presentation purpose, we have included  %%
%% \bigskip command. Please ignore this.       %%
%%=============================================%%

\backmatter

%\bmhead{Supplementary information}

%If your article has accompanying supplementary file/s please state so here. 

%Authors reporting data from electrophoretic gels and blots should supply the full unprocessed scans for key as part of their Supplementary information. This may be requested by the editorial team/s if it is missing.

%Please refer to Journal-level guidance for any specific requirements.

\bmhead{Acknowledgements}

This work is supported by National Science and Technology Major Project (2022ZD0116800), Program of National Natural Science Foundation of China (12425114, 62141605, 12201026, 12301305, 62441617, 12501702), the Fundamental Research Funds for the Central Universities, Beijing Natural Science Foundation (Z230001), the Opening Project of the State Key Laboratory of General Artificial Intelligence(Project No. SKLAGI2025OP16), and Bejing Advanced Innovation Center for Future Blockchain and Privacy Computing.

%\section*{Declarations}

%Some journals require declarations to be submitted in a standardised format. Please check the Instructions for Authors of the journal to which you are submitting to see if you need to complete this section. If yes, your manuscript must contain the following sections under the heading `Declarations':

%\begin{itemize}
%\item Conflict of interest/Competing interests (check journal-specific guidelines for which heading to use)
%\item Ethics approval and consent to participate
%\item Consent for publication
%\item Data availability 
%\end{itemize}

%\noindent
%If any of the sections are not relevant to your manuscript, please include the heading and write `Not applicable' for that section. 

\bigskip

%\begin{appendices}

%\section{Section title of first appendix}\label{secA1}




% \end{appendices}


\bibliography{ref}% common bib file

\end{document}

\title[Article Title]{\textbf{Supporting Information for}\par{Indirect Reciprocity with Environmental Feedback}}


\author[1,4,6]{\fnm{Yishen} \sur{Jiang}}
\author*[2,4,6,7,9]{\fnm{Xin} \sur{Wang}}\email{wangxin\_1993@buaa.edu.cn}
\author[1,4,6]{\fnm{Ming} \sur{Wei}}
\author[2,4,6]{\fnm{Wenqiang} \sur{Zhu}}
\author[2,4,6,7]{\fnm{Longzhao} \sur{Liu}}
\author[10]{\fnm{Hongwei} \sur{Zheng}}
\author*[2,3,4,5,6,7,8]{\fnm{Shaoting} \sur{Tang}}\email{tangshaoting@buaa.edu.cn}

\affil[1]{School of Mathematical Sciences, Beihang University, Beijing 100191, China}
\affil[2]{School of Artificial Intelligence, Beihang University, Beijing 100191, China}
\affil[3]{Hangzhou International Innovation Institute, Beihang University, Hangzhou 311115, China}
\affil[4]{Key Laboratory of Mathematics, Informatics and Behavioral Semantics, Beihang University, Beijing 100191, China}
\affil[5]{Institute of Medical Artificial Intelligence, Binzhou Medical University, Yantai 264003, China}
\affil[6]{Zhongguancun Laboratory, Beijing 100094, China}
\affil[7]{Beijing Advanced Innovation Center for Future Blockchain and Privacy Computing, Beihang University, Beijing 100191, China}
\affil[8]{Institute of Trustworthy Artificial Intelligence, Zhejiang Normal University, Hangzhou, 310013}
\affil[9]{State Key Laboratory of General Artificial Intelligence, BIGAI, Beijing, China}
\affil[10]{Beijing Academy of Blockchain and Edge Computing, Beijing 100085, China}



\maketitle

This Supplementary Information supports the main text with a comprehensive analysis into five sections. Section 1 reviews the mathematical formulation of the reputation assessment mechanism and defines the baseline probabilities of acquiring a good reputation based on donor intent and recipient standing. Building on this foundation, Section 2 performs a stability analysis in static environments to derive the precise environmental thresholds that determine the stability of the Discriminator strategy against unconditional strategies. We then extend the framework to dynamic environments in Section 3 and investigate the coevolutionary dynamics of organized pairwise strategy competitions. Section 4 expands the analysis to the three-strategy system. We first add the hybrid evolutionary nature of Simple Standing and Shunning in static environments. We then explore the dynamics under environmental feedback and examine the resilience of these norms against reduced environmental sensitivity. Finally, Section 5 investigates the evolution of social norms under dynamic feedback to elucidate how the environmental feedback rate regulates evolutionary timescales and how environmental sensitivity drives critical bifurcations in norm competition.

\section{Reputation assessment}\label{sec1}

Here, we review the mathematical formulation of reputation presented in the main text. First, we define the four probabilities of obtaining a good reputation when interacting with others: 
\begin{itemize}
    \item interacting with a person who has a good reputation and intending to cooperate, $P_{GC}=(1-u_c)(1-u_a)+u_c u_a=\epsilon$;
    \item interacting with a person with a good reputation and intending to defect, $P_{GD}=u_a$;
    \item interacting with a person with a bad reputation and intending to cooperate, $P_{BC}=p(\epsilon-u_a)+q(1-\epsilon-u_a)+u_a$;
    \item interacting with a person with a bad reputation and intending to defect, $P_{BD}=q(1-2u_a)+u_a$.
\end{itemize}
Here, $p$ and $q$ represent the probabilities of cooperating with individuals possessing good and bad reputations, respectively, the specific meanings of which have been elaborated in the main text.

Next, we calculate the probability $r^S_{j,i}$ that strategy $S$ in group $i$ is perceived as having a good reputation by group $j$, where $S \in \{ALLC, ALLD, DISC\}$. 
\begin{itemize}
    \item For ALLC, $r^{ALLC}_{j,i}=r_{j, \cdot} P_{GC}+(1- r_{j, \cdot}) P_{BC}$.
    \item Similarly, $r^{ALLD}_{j,i}=r_{j, \cdot} P_{GD}+(1- r_{j, \cdot}) P_{BD}$.
    \item For DISC, the situation is more complex: $r^{DISC}_{j,i}=\delta_{j,i} [r_{j, \cdot} P_{GC}+(1- r_{j, \cdot})P_{BD}]+(1-\delta_{j,i})[R_{j,i} P_{GC}+(r_{j, \cdot}-R_{j,i})P_{GD}+(r_{i, \cdot}-R_{j,i})P_{BC}+(1- r_{j, \cdot}- r_{i, \cdot}+R_{j,i})P_{BD}]$, where $R_{j,i}=\sum_{k} \nu_{k} \sum_{S} f^S_k r^S_{i,k} r^S_{j,k}$.
\end{itemize}   
The average reputation of each strategy and the population average reputation are given by $r^S=\sum_{i}\sum_{j}{\nu_i\nu_jr_{j,i}^S}$ and $r=\sum_{i}\sum_{j}{\nu_i\nu_jr_{j,i}}$, respectively.

\section{Invasibility of discriminators in a single group}\label{sec2}

We begin by reviewing the scenario presented in the main text where $K=1$, meaning the entire population adheres to a single social norm. Under this setting, we examine the ability of the Discriminator strategy to invade other strategies and its ability to resist invasion.

\subsection{ALLD and DISC}

The condition for a population of pure Defectors (ALLD) to resist invasion by a small number of Discriminators (DISC) is given by the following linear stability analysis:
\begin{equation*}
    \left.\frac{\partial\dot{f^{DISC}}}{\partial f^{DISC}}\right|_{f^{DISC}=0}<0.
\end{equation*}
Letting $f=f^{DISC}$, and according to the replicator dynamic equation, the above condition is equivalent to:
\begin{align*}
\partial_f\left[f\left(\pi^{DISC}-\bar{\pi}\right)\right|_{f=0}&<0 \\
\partial_f\left[\left(f-f^2\right)\left(\pi^{DISC}-\pi^{ALLD}\right)\right|_{f=0}&<0\\
\left.\left[\left(1-2f\right)\left(\pi^{DISC}-\pi^{ALLD}\right)+f\left(1-f\right)\partial_f\left(\pi^{DISC}-\pi^{ALLD}\right)\right]\right|_{f=0}&<0 \\
\left.\pi^{DISC}\right|_{f=0}&<\left.\pi^{ALLD}\right|_{f=0}\\
\left.[bfr^{DISC}-nrc-(1-n)(1-r)c]\right|_{f=0} &< \left.[bfr^{ALLD}-(1-n)c]\right|_{f=0}\\
cr\left(2n-1\right)&>0\\
n&>\frac{1}{2}
\end{align*}

\setcounter{figure}{0}
\renewcommand{\thefigure}{S\arabic{figure}}

Similarly, the condition for discriminators to resist invasion by defectors is:
\begin{equation*}
    \left.\frac{\partial\dot{f^{ALLD}}}{\partial f^{ALLD}}\right|_{f^{ALLD}=0}<0.
\end{equation*}
Note that $f^{ALLD}=1-f^{DISC}=1-f$, so stability at $f^{ALLD}=0$ is equivalent to stability at $f=1$. Similarly, this requires the payoff of ALLD when rare to be lower than the payoff of DISC when resident:
\begin{align*}
\left.\pi^{ALLD}\right|_{f=1}&<\left.\pi^{DISC}\right|_{f=1}\\
\left.\left[bfr^{ALLD}-\left(1-n\right)c\right]\right|_{f=1}&<\left.\left[bfr^{DISC}-nrc-\left(1-n\right)\left(1-r\right)c\right]\right|_{f=1}\\
\left.\left[bf\left(r^{ALLD}-r^{DISC}\right)\right]\right|_{f=1}&<\left.\left[c\left(1-2n\right)r\right]\right|_{f=1}\\
\left.\left[b\left(r^{DISC}-r^{ALLD}\right)\right]\right|_{f=1}&>\left.\left[c\left(2n-1\right)r\right]\right|_{f=1}
\end{align*}
Since $r^{DISC}=rP_{GC}+\left(1-r\right)P_{BD}$ and $r^{ALLD}=rP_{GD}+\left(1-r\right)P_{BD}$, we have
\begin{align*}
br\left(P_{GC}-P_{GD}\right)&>c\left(2n-1\right)\\
\frac{b}{c}&>\frac{2n-1}{\epsilon-u_a}
\end{align*}
This inequality can also be written in terms of the environmental parameter $n$:
\begin{equation*}
    n<\frac{1+\frac{b}{c}(\epsilon-u_a)}{2}
\end{equation*}

In addition to the two boundary equilibria at $f^{DISC}=1$ and $f^{DISC}=0$, a third equilibrium exists in the interior:
\begin{align*}
\dot{f^{DISC}}&=0\\
\pi^{DISC}&=\pi^{ALLD}\\
f_{ALLD}^{DISC}&=\frac{c}{b}\frac{2n-1}{\epsilon-u_a}
\end{align*}
We obtained above the condition for DISC and ALLD to simultaneously mutually resist invasion is $\frac{1}{2}<n<\frac{1+(b/c)(\epsilon-u_a)}{2}$. It is worth noting that this range coincides exactly with the domain where the internal equilibrium $f_{ALLD}^{DISC}$ exists (i.e., $0 < f_{ALLD}^{DISC} < 1$). Therefore, when the environmental state $n$ is within this range, the boundary states $f^{DISC}=1$ and $f^{DISC}=0$ are evolutionarily stable, while the internal equilibrium is unstable.

\subsection{ALLC and DISC}

Next, we examine the mutual invasibility between a population of pure cooperators and a population of discriminators. Similar to the derivation above, the condition for cooperators to resist invasion by discriminators (again letting $f^{DISC}=f$) is:
\begin{align*}
    \left.\frac{\partial\dot{f^{DISC}}}{\partial f^{DISC}}\right|_{f^{DISC}=0}&<0\\
    \partial_f\left[f\left(\pi^{DISC}-\bar{\pi}\right)\right|_{f=0}&<0\\
    \left.\pi^{DISC}\right|_{f=0}&<\left.\pi^{ALLC}\right|_{f=0}\\
    \left.\left[b\left(f^{ALLC}+fr^{DISC}\right)-nrc-\left(1-n\right)\left(1-r\right)c\right]\right|_{f=0}&<\left.\left[b\left(f^{ALLC}+fr^{ALLC}\right)-nc\right]\right|_{f=0}\\
    \left(2n-1\right)\left(1-r\right)&<0\\
    n&<\frac{1}{2}
\end{align*}
This means that the ALLC strategy is evolutionarily stable when the environmental state satisfies $n < 1/2$. Conversely, when the DISC strategy is dominant, the condition for it to resist invasion by rare ALLC strategies is:
\begin{equation*}
    \left.\frac{\partial\dot{f^{ALLC}}}{\partial f^{ALLC}}\right|_{f^{ALLC}=0}<0.
\end{equation*}
Noting that stability at $f^{ALLC}=0$ is equivalent to stability at $f=1$, we have:
\begin{align*}
    \left.\pi^{ALLC}\right|_{f=1}&<\left.\pi^{DISC}\right|_{f=1}\\
    [b(f^{ALLC}+fr^{ALLC})-nc]_{f=1} &< [b(f^{ALLC}+fr^{DISC})-nrc-(1-n)(1-r)c]_{f=1}\\
    \left.b\left(r^{ALLC}-r^{DISC}\right)\right|_{f=1}&<c\left(1-r\right)\left(2n-1\right)\\
    b\left(P_{BC}-P_{BD}\right)&<c\left(2n-1\right)
\end{align*}
This inequality can be rewritten in terms of the environmental parameter $n$:
\begin{equation*}
    n>\frac{1+\frac{b}{c}\left(P_{BC}-P_{BD}\right)}{2}
\end{equation*}
Based on the sign and value of $P_{BC}-P_{BD}$ under different social norms ($P_{BC}-P_{BD}$ is 0 under SH and SS, $u_a-\epsilon < 0$ under SJ, and $\epsilon-u_a > 0$ under SC), the above condition can be further refined as:
\begin{align*}
\begin{cases}
n>\frac{1}{2}, & \text{under SH and SS}, \\
\frac{b}{c}>\frac{2n-1}{u_a-\epsilon}, & \text{under SJ}, \\
\frac{b}{c}<\frac{2n-1}{\epsilon-u_a}, & \text{under SC}.
\end{cases}
\end{align*}
In addition to the boundary equilibria, by solving $\dot{f^{DISC}}=0$ (i.e., $\pi^{DISC}=\pi^{ALLC}$), we can obtain a potential internal equilibrium:
\begin{equation*}
    f_{ALLC}^{DISC}=\frac{c}{b}\frac{2n-1}{P_{BC}-P_{BD}}.
\end{equation*}
This equilibrium does not exist under SH and SS norms; it exists and is unstable under SJ when $n < 1/2$; and it exists and is stable under SC when $n > 1/2$.

\section{Competition between two strategies under dynamic environments}

When examining the competition between ALLC and ALLD, the reputation mechanism becomes irrelevant because the actions of these strategies are independent of the opponent's reputation. Under these conditions, the system exhibits characteristic periodic oscillatory dynamics, a typical manifestation of the "tragedy of the commons" driven by environmental feedback. Our simulation results successfully replicate the findings of classic environmental feedback models (Fig.~\ref{figS3}): the environmental feedback rate, $\eta$, determines the evolutionary time scale of the system, whereas the parameter $\theta$ dictates the position of the equilibrium point.

Regarding the competitive dynamics between ALLD and DISC, simulation results validate the analysis presented in the main text: the evolutionary trajectory of the system is primarily governed by the combined parameter $q$, where identical $q$ values yield essentially consistent evolutionary dynamics. The system typically converges to a unique globally stable state dominated by the DISC strategy. However, a notable dynamical phenomenon occurs under conditions of an extremely low benefit parameter $b$, where the system no longer stabilizes at a single equilibrium point (Fig.~\ref{figS4}E), but instead gives rise to complex oscillations characterized by a heteroclinic cycle. Furthermore, concerning the influence of environmental feedback parameters, increasing parameter $\theta$ contributes to improving the final level of environmental recovery (Fig.~\ref{figS4}B-D), whereas the environmental feedback rate $\eta$, while regulating the time scale (speed) of the evolutionary process, has no qualitative impact on the evolutionary outcomes of the system.

In the competition between ALLC and DISC, different social norms significantly shape the long-term evolutionary outcomes of the system, resulting in distinct dynamical behaviors. Specifically, under SJ and SS norms, the system exhibits consistent evolutionary trends; the DISC strategy demonstrates strong robustness, ultimately emerging as the sole globally evolutionarily stable state (Fig.~\ref{figS5}A-F). In contrast, the SC norm fosters strategic diversity, permitting the formation of a stable coexistence state between ALLC and DISC (Fig.~\ref{figS5}G-I). The situation under the SH norm is the most complex, with evolutionary outcomes being highly dependent on $\theta$: when the value of $\theta$ is small, the system may lose stability, giving rise to periodic oscillatory behaviors (Fig.~\ref{figS5}J); however, as $\theta$ increases, the DISC strategy regains dominance and stabilizes the system (Fig.~\ref{figS5}L).

\section{Evolution of three strategies}

We first extends the analysis to Simple Standing and Shunning in static environments, complementing the results for Stern Judging and Scoring presented in the main text (Fig.~\ref{figS6}). The evolutionary outcomes under these two norms exhibit hybrid characteristics governed by the environmental state $n$. In resource-poor environments where $n < 1/2$, the system is globally dominated by ALLC. This pattern mirrors the dynamics observed under Scoring. Conversely, when the environment becomes abundant with $n > 1/2$, the dynamics shift to a bistable configuration between ALLD and DISC. This resembles the stability found under Stern Judging. Therefore, SS and SH act as intermediate norms. They fail to sustain discriminators against unconditional cooperators under scarcity like Scoring, yet successfully support the segregation of discriminators from defectors under abundance like Stern Judging, thereby avoiding the boundary coexistence seen in Scoring.

While the main text has detailed the evolutionary dynamics of the system under the SJ norm, we supplement these findings by presenting a comparison of evolutionary outcomes governed by three other key social norms under strictly identical parameter settings. Under SS and SH norms, the system exhibits consistent evolutionary outcomes, ultimately converging to a unique globally stable state dominated by the DISC strategy (Fig.~\ref{figS7}A and Fig.~\ref{figS7}D). In contrast, the evolutionary trajectory under the SC norm is more complex and intricate; the system enters a prolonged pseudo-steady state phase before eventually reaching a bistable configuration (Fig.~\ref{figS7}G). Furthermore, we employed Monte Carlo individual-based simulations to cross-validate the accuracy of the aforementioned deterministic dynamical results (Fig.~\ref{figS7}B-C and Fig.~\ref{figS7}E-F). It should be specifically noted that for the SC norm, due to its complex phase space structure characterized by the coexistence of multiple stable attractors and significant pseudo-steady state regions, stochastic simulation processes are highly prone to jumping between different basins of attraction, making the results difficult to present clearly. Therefore, for the sake of conciseness in presentation, we have omitted the Monte Carlo simulation illustrations corresponding to the SC norm here.

Next, we shift our focus to exploring how the environmental feedback parameter $\theta$ reshapes the evolutionary dynamics of the system. In general, reducing the value of $\theta$ significantly suppresses the recovery capability of the environmental state. However, different social norms exhibit distinct responses to this. Under SJ and SS norms, due to their intrinsic mechanisms maintaining high levels of population cooperation, the evolutionary stable structure of the system shows strong robustness to the weakening of environmental resilience and remains largely unaffected (Fig.~\ref{figS8}A-B and Fig.~\ref{figS8}E-F). In contrast, the SH norm is more sensitive to changes in $\theta$: when $\theta$ is small, the system loses stability and instead exhibits periodic oscillatory evolutionary behaviors (Fig.~\ref{figS8}D and Fig.~\ref{figS8}H). The SC norm, meanwhile, presents a complex attractor landscape; as $\theta$ decreases, multiple internal stable equilibrium points (coexistence states of all three strategies) emerge within the phase space, allowing the environmental state $n$ to stabilize at certain intermediate levels rather than tending towards extremes (Fig.~\ref{figS8}C and Fig.~\ref{figS8}G).

\section{Evolution of norms under dynamic environments}

This section primarily illustrates how two key parameters, the environmental feedback rate $\eta$ and environmental sensitivity $\theta$, jointly shape the evolutionary outcomes of competition among social norms. First, consistent with previous findings at the strategy level, the environmental feedback rate $\eta$ does not qualitatively change the dynamics of norm competition. Instead, it mainly regulates the time scale of evolution, influencing how quickly the system converges to its eventual equilibrium, while leaving the type and location of that equilibrium unchanged.

In contrast, environmental sensitivity $\theta$ exhibits significant shaping power over the landscape of norm competition. In the interplay between SJ and SH, as the value of $\theta$ increases, the system undergoes a critical dynamical transition (bifurcation): evolving from an initial singular globally stable state dominated by the SJ norm into a bistable configuration where SJ and SH can coexist long-term. Notably, across this entire range of parameter variation, regardless of which norm is dominant, the system maintains the environment in its most abundant state (Fig.~\ref{figS9}).

Conversely, in the competitive scenario between SC and SH, the system consistently exhibits bistability within the examined range of $\theta$ parameters. However, increasing $\theta$ fundamentally alters the environmental quality associated with these stable states: under low $\theta$ conditions, norm competition causes the environment to fall into an impoverished state, whereas higher $\theta$ values successfully facilitate a transformation of the environment from impoverished to abundant, achieving ecological recovery (Fig.~\ref{figS10}).

% Figure

\begin{figure}[p]
\centering
\includegraphics[width=1\textwidth]{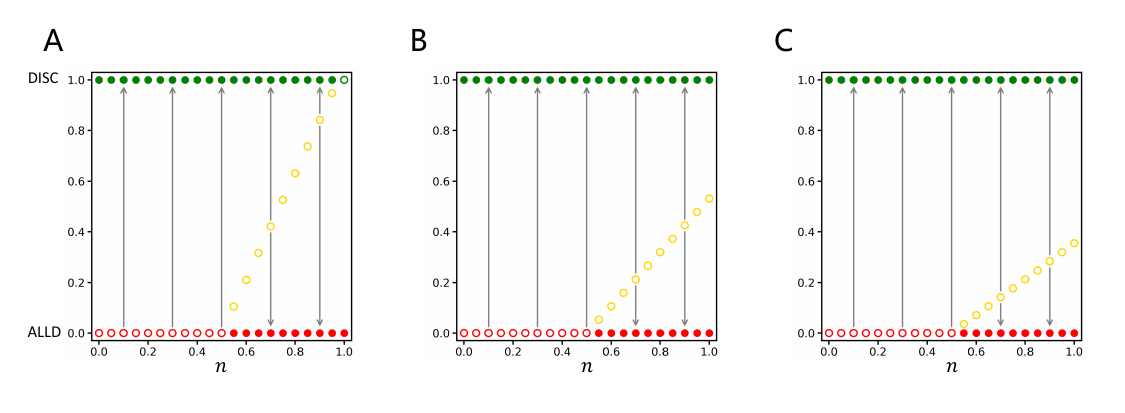}
\caption{\textbf{Competition between ALLD and DISC in static environments.} Solid circles indicate evolutionarily stable equilibria, while hollow circles indicate unstable equilibria (green for DISC, red for ALLD, and yellow for internal coexistence). Grey arrows indicate the flow of evolutionary dynamics of strategy frequencies. (A) $b=1.01$. This panel shows that under low payoff conditions, as the environment $n$ deteriorates, the system sequentially transitions through regions of DISC stability, bistability, and finally, ALLD stability. (B) $b=2$ and (C) $b=3$. These panels illustrate scenarios with higher payoff conditions. Compared to (A), higher $b$ values eliminate the region of sole ALLD stability at high $n$ values, maintaining the system in a bistable state even in harsh environments. With further increases in $b$, the basin of attraction for the DISC strategy expands.}\label{figS1}
\end{figure}

\begin{figure}[p]
\centering
\includegraphics[width=1\textwidth]{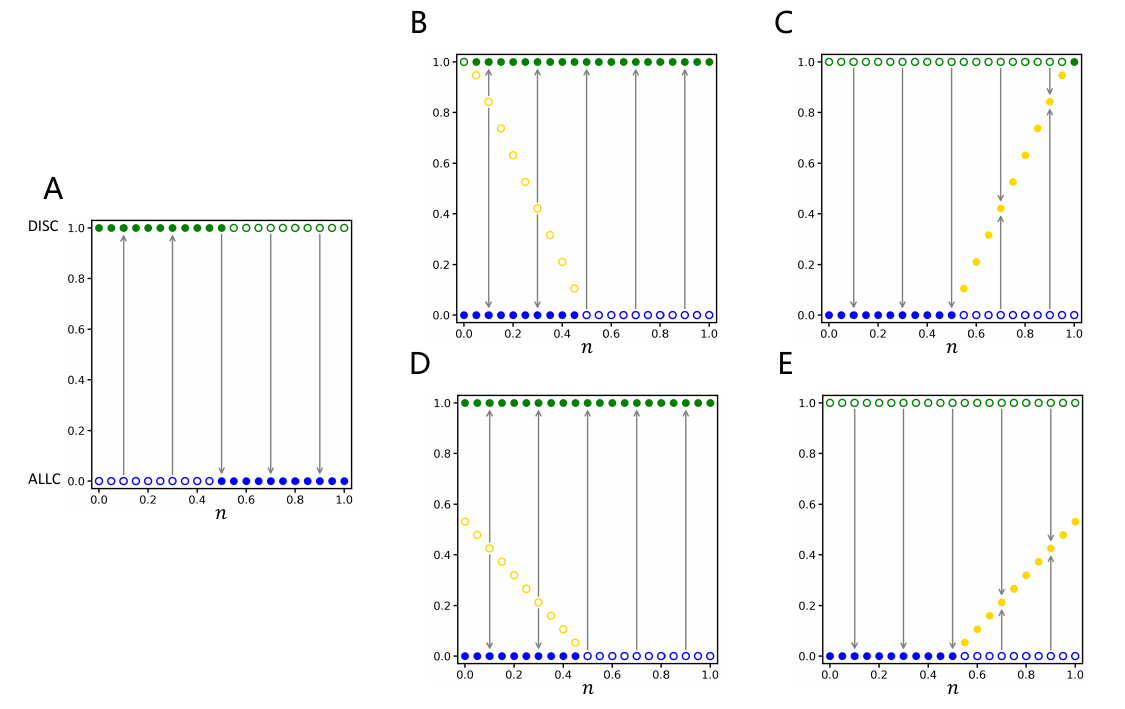}
\caption{\textbf{Competition between ALLC and DISC in static environments.} Solid circles indicate evolutionarily stable equilibria, while hollow circles indicate unstable equilibria (green for DISC, blue for ALLC, and yellow for internal coexistence). Grey arrows show the flow of evolutionary dynamics. (A) SH and SS norms. Stability switches at n=0.5. (B) SJ norm and (C) SC norm, with b=1.01. Under low payoff conditions, both SJ and SC exhibit three distinct regions of stability. (D) SJ norm and (E) SC norm, with b=2. At higher payoffs, the region of sole ALLC stability disappears under the SJ norm, whereas the region of sole DISC stability disappears under the SC norm.}\label{figS2}
\end{figure}

\begin{figure}[p]
\centering
\includegraphics[width=1\textwidth]{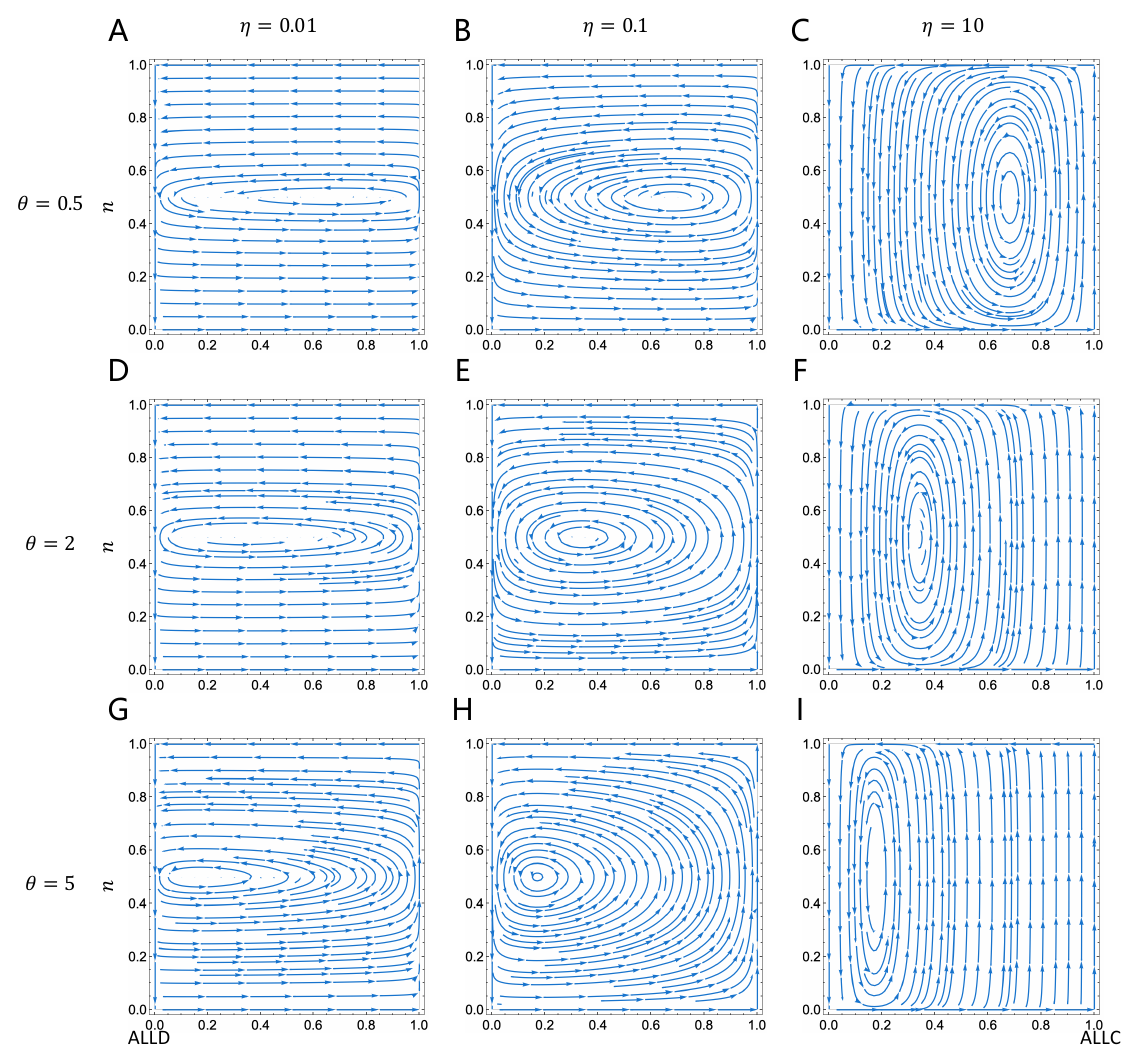}
\caption{\textbf{Coevolution of ALLC and ALLD in dynamic environments.} The horizontal axis represents the proportion of the ALLC strategy in the population, and the vertical axis represents the environmental state $n$. Columns represent different $\eta$ values (0.01, 0.1, 10), and rows represent different $\theta$ values (0.5, 2, 5). These phase diagrams show the periodic oscillatory dynamics exhibited by the system in the absence of reputation mechanisms.}\label{figS3}
\end{figure}

\begin{figure}[p]
\centering
\includegraphics[width=1\textwidth]{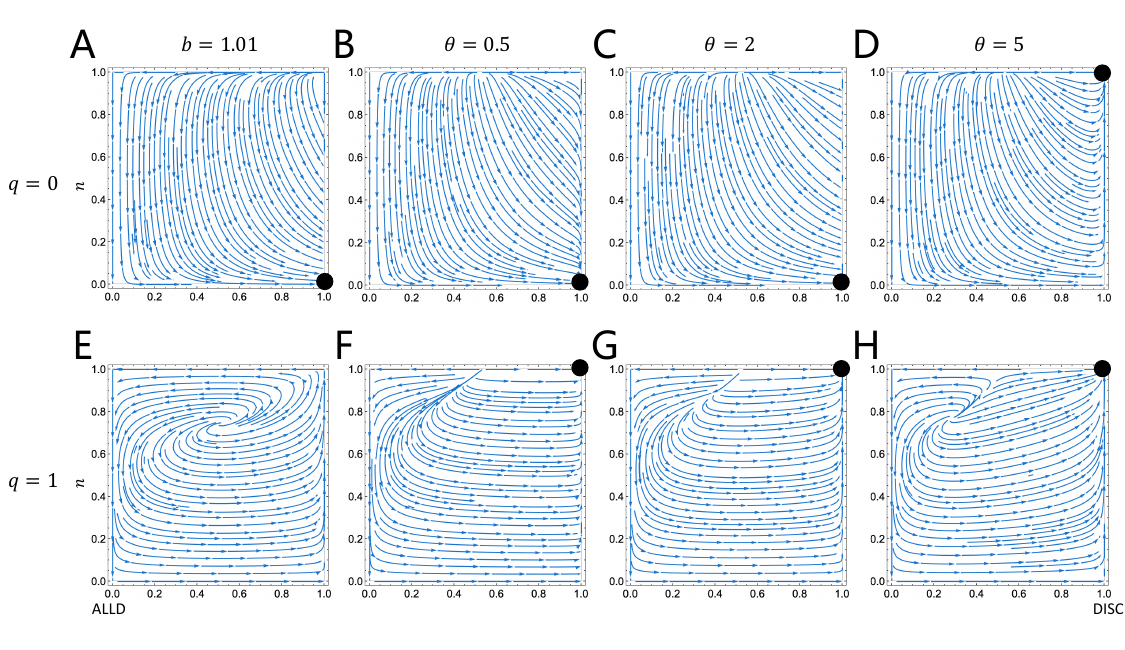}
\caption{\textbf{Coevolution of ALLD and DISC in a dynamic environment.} The horizontal axis represents the proportion of the DISC strategy in the population, and the vertical axis represents the environmental state $n$. Solid black circles indicate stable equilibria. The first row shows evolutionary results under SC and SH norms ($q=0$), and the second row shows results under SJ and SS norms ($q=1$). Parameters: $\eta=0.1$; for (A) and (E), $b=1.01$; for (B)-(D) and (F)-(H), $b=2$.}\label{figS4}
\end{figure}

\begin{figure}[p]
\centering
\includegraphics[width=1\textwidth]{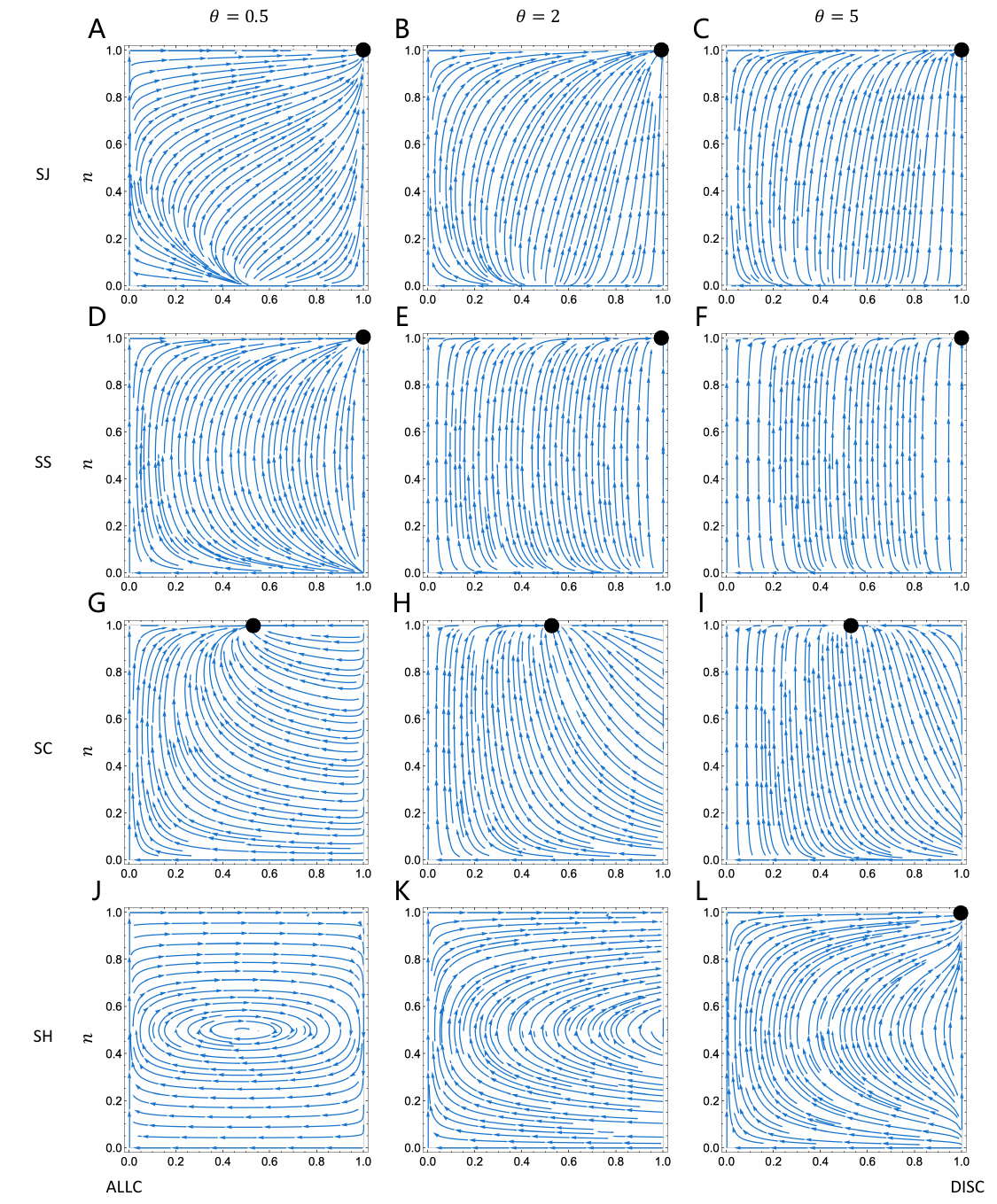}
\caption{\textbf{Coevolution of ALLC and DISC in a dynamic environment.} The horizontal axis represents the proportion of the DISC strategy in the population, and the vertical axis represents the environmental state $n$. Solid black circles indicate stable equilibria. Different rows correspond to different social norms, and different columns correspond to different $\theta$ values. Parameters: $\eta=0.1$, $b=2$.}\label{figS5}
\end{figure}

\begin{figure}[p]
\centering
\includegraphics[width=1\textwidth]{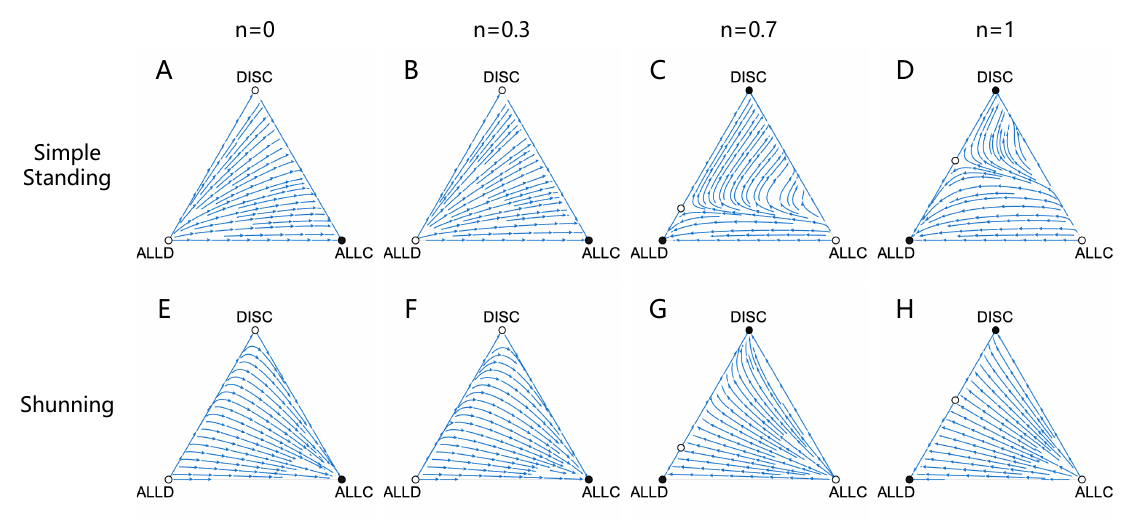}
\caption{\textbf{Evolutionary dynamics of three strategies in static environments under SS and SH norms.} The simplexes illustrate evolutionary trajectories under two social norms (rows) and varying static environments ($n$, columns). Solid circles denote stable equilibria, and open circles denote unstable equilibria. Under Simple Standing and Shunning, ALLC is the globally stable equilibrium in poor environments, whereas rich environments support bistability between ALLD and DISC. Parameter: b=2}\label{figS6}
\end{figure}

\begin{figure}[p]
\centering
\includegraphics[width=1\textwidth]{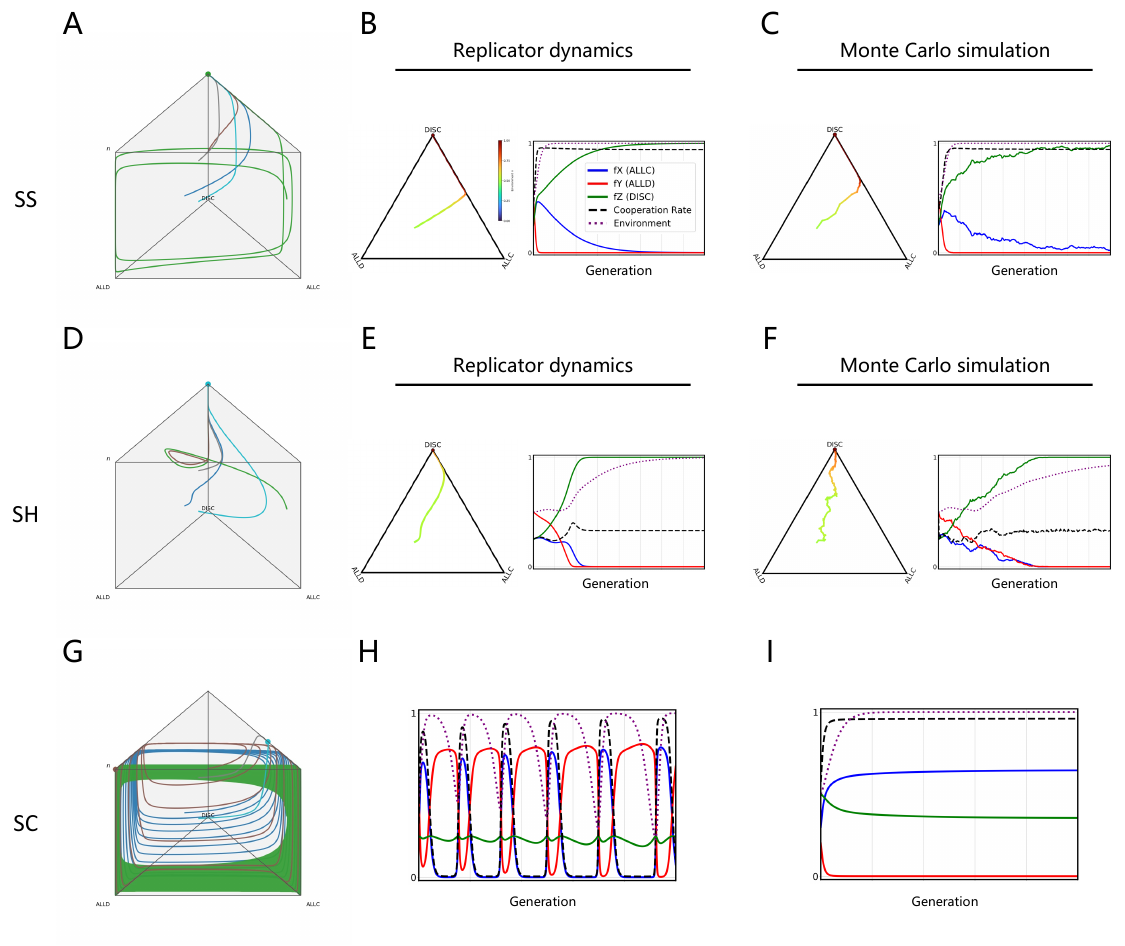}
\caption{\textbf{Coevolution of three strategies and dynamic environments.} he rows in the figure correspond to different social norms: SS (top row), SH (middle row), and SC (bottom row). (A, D, G) display the overall evolutionary trajectories of strategy frequencies within the three-dimensional simplex. (B, E) and (C, F) compare the temporal evolution of deterministic replicator dynamics with corresponding stochastic Monte Carlo individual-based simulation results for selected initial points, validating the predictive capability of the theoretical model. (H, I) specifically highlight the complex dynamical characteristics under the SC norm: due to the existence of distinct basins of attraction, different initial conditions can lead to either long-term pseudo-steady state oscillations (H) or eventually reach a stable coexistence state (I). Parameters: $b=3, \theta=3, \eta=0.1$.}\label{figS7}
\end{figure}

\begin{figure}[p]
\centering
\includegraphics[width=1\textwidth]{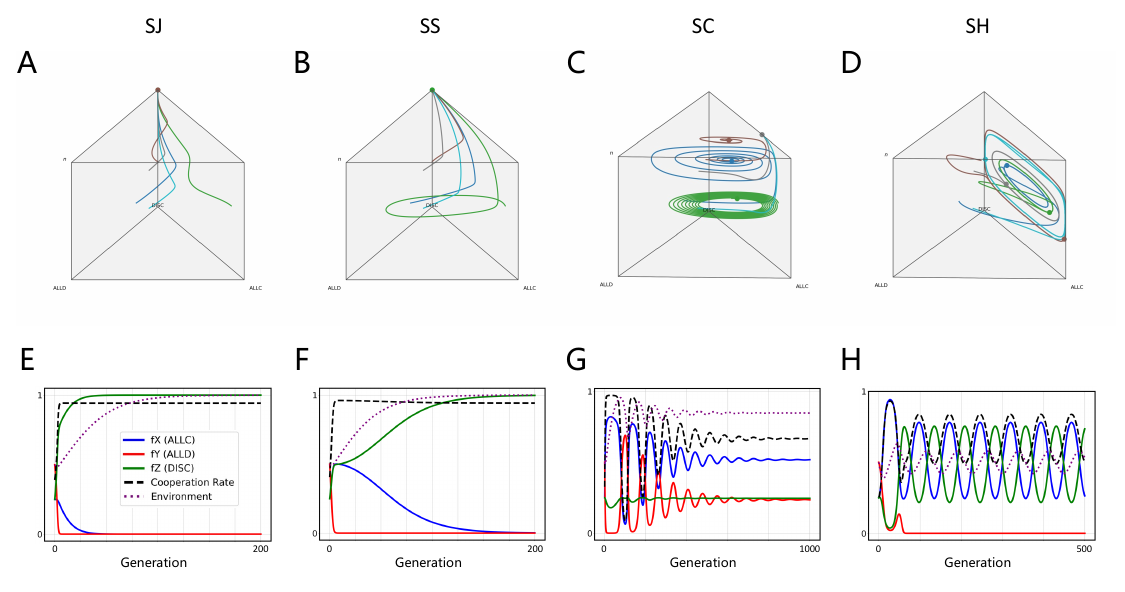}
\caption{\textbf{Coevolutionary dynamics of different social norms under low environmental feedback sensitivity.} (A-D) show evolutionary trajectories in the three-dimensional strategy space. (E-H) show time evolution processes from the specific initial point. Each column corresponds to SJ, SS, SC, and SH norms, respectively. SJ and SS norms exhibit a DISC stable state, the SH norm exhibits oscillatory dynamics, and the SC norm exhibits an internal coexistence stable state. Parameters: $b=3, \theta=0.5, \eta=0.1$.}\label{figS8}
\end{figure}

\begin{figure}[p]
\centering
\includegraphics[width=1\textwidth]{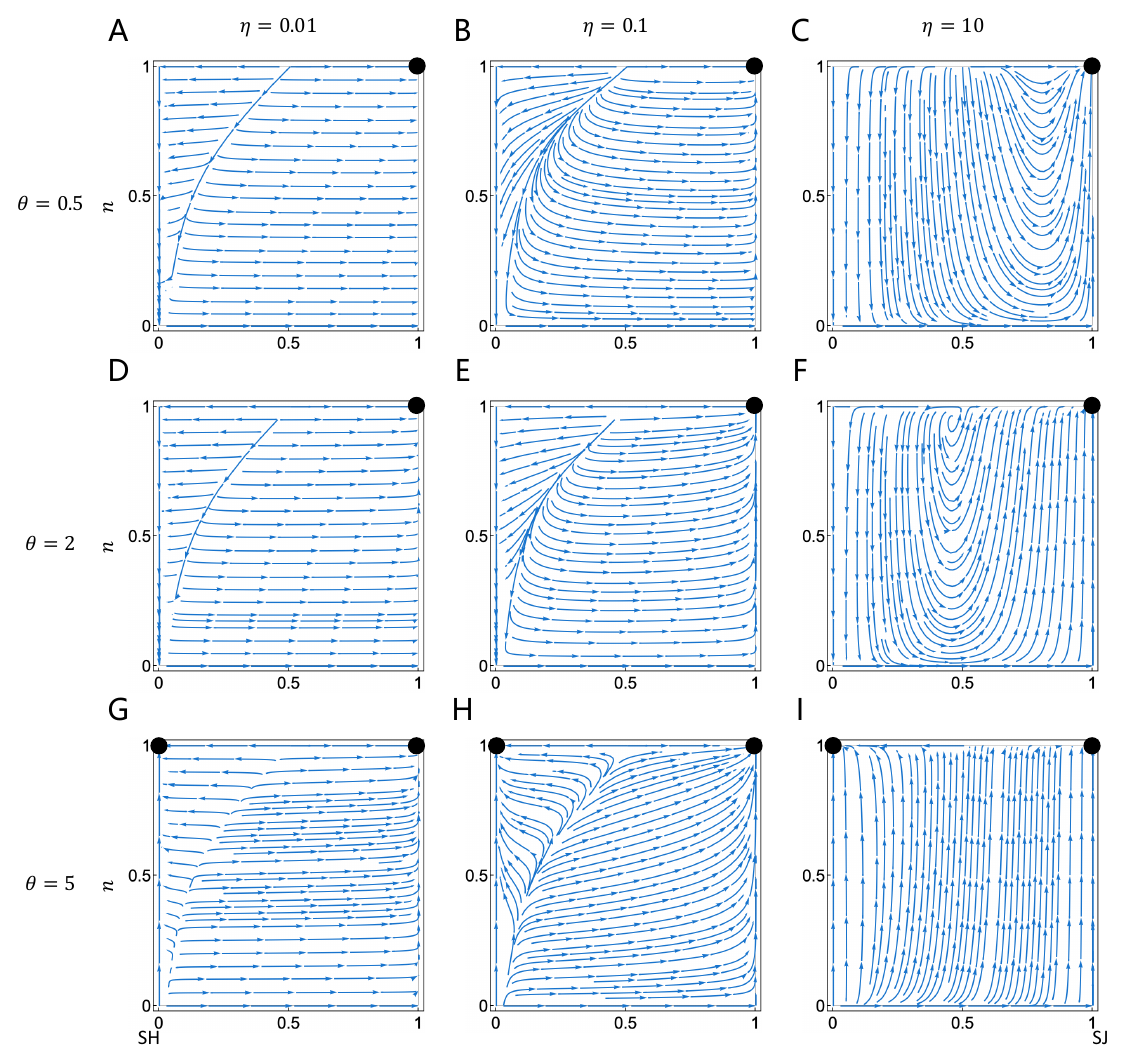}
\caption{\textbf{Phase diagrams of the coevolutionary dynamics of SJ and SH norms a dynamic environments.} The horizontal axis represents the proportion of the SJ norm in the population, and the vertical axis represents the environmental state $n$. Solid black circles indicate evolutionarily stable equilibria. Different columns represent different environmental feedback rates $\eta$ (0.01, 0.1, 10), and different rows represent different environmental sensitivities $\theta$ (0.5, 2, 5). As $\theta$ increases, the system evolves from SJ monostability to bistability of SJ and SH.} \label{figS9}
\end{figure}

\begin{figure}[p]
\centering
\includegraphics[width=1\textwidth]{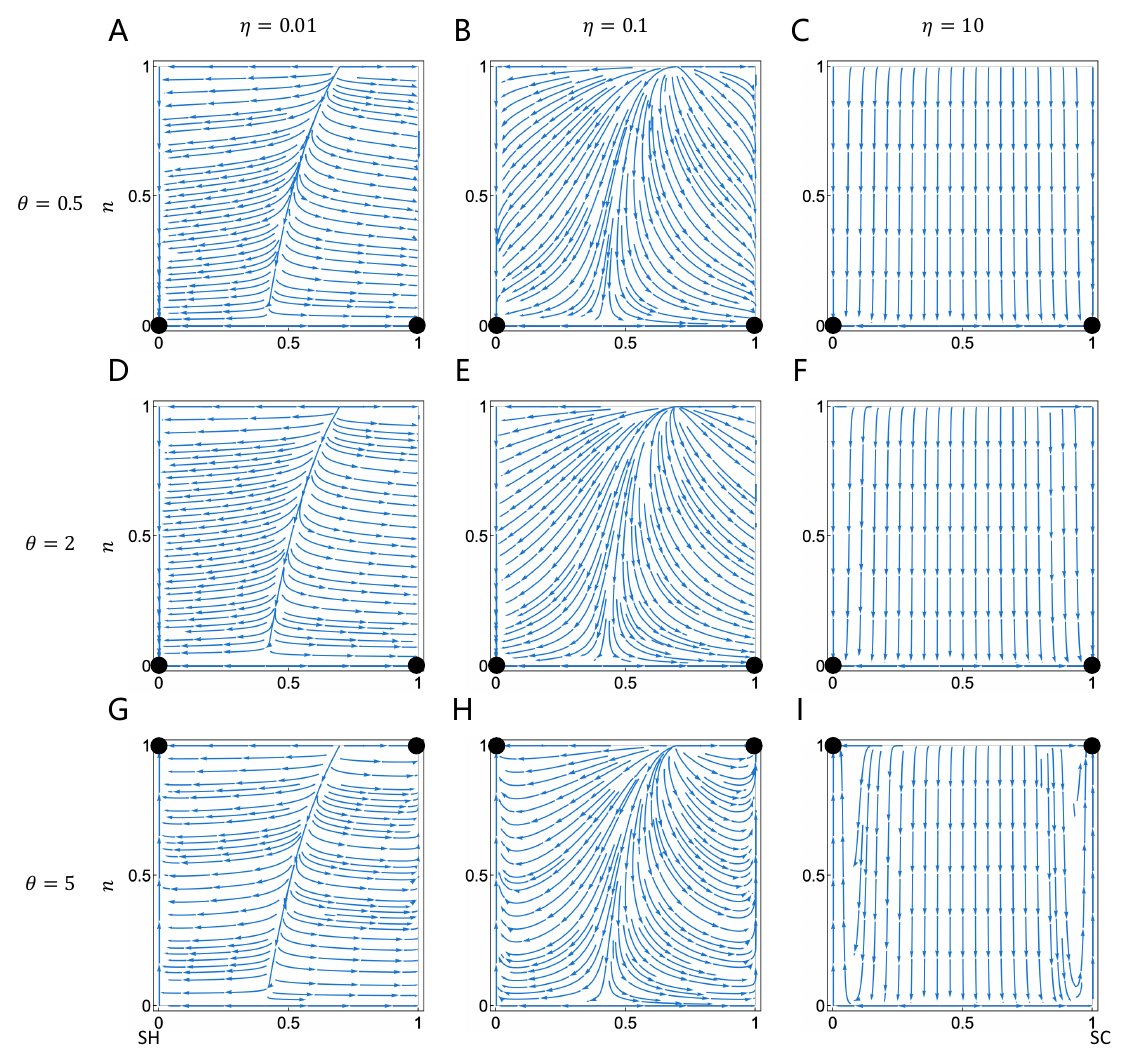}
\caption{\textbf{Phase diagrams of the coevolutionary dynamics of SC and SH norms in dynamic environments.} The horizontal axis represents the proportion of the SC norm in the population, and the vertical axis represents the environmental state $n$. Solid black circles indicate evolutionarily stable equilibria. Different columns represent different environmental feedback rates, and different rows represent different environmental sensitivities. Within the parameter range shown, the system consistently exhibits a bistable configuration. As $\theta$ increases, the environment transitions from impoverished to abundant. Parameters: $b=2, K=2$.}\label{figS10}
\end{figure}

\backmatter

\bigskip

%\begin{appendices}

%\section{Section title of first appendix}\label{secA1}




% \end{appendices}


%\bibliography{ref}% common bib file